\documentclass[11pt]{article}
\usepackage[lmargin=1in,bmargin=.8in,tmargin=1in,rmargin=1in]{geometry}
\usepackage{amsmath,amssymb,amsfonts,amsthm,mathtools,mathrsfs}
\usepackage{graphicx}
\usepackage{float}
\usepackage{bm}
\usepackage{csquotes}

\usepackage{epsfig}
\usepackage{amsmath,amsthm,mathtools,mathrsfs}
\usepackage{ulem}
\usepackage{amssymb}
\usepackage[dvipsnames]{xcolor}
\usepackage{amsfonts}
\usepackage{physics}
\usepackage{relsize}

\usepackage[utf8]{inputenc}
\usepackage{float}

\usepackage{xcolor, soul}
\usepackage{colortbl}
\usepackage{comment}
\usepackage{mdframed}

\usepackage{subcaption}
\usepackage[shortlabels]{enumitem}
\usepackage{tikz}
\usepackage{authblk}
\usepackage[numbers,sort&compress]{natbib}
\usepackage{hyperref}
\usepackage{cleveref}

\newcommand{\be}{\begin{equation}}
\newcommand{\ee}{\end{equation}}
\newcommand{\bea}{\begin{eqnarray}}
\newcommand{\eea}{\end{eqnarray}}

\begin{document}
\title{Genotype specificity and spatial arrangement govern the direction and magnitude of selection in variable environments}
\author[1,$\star$]{Hossein Nemati}
\author[2,$\star$,$\dagger$]{Kamran Kaveh}
\author[3]{Jakub Svoboda}
\author[4]{Mohammad Reza Ejtehadi}
\author[5]{Krishnendu Chatterjee}

\affil[1]{Utrecht University, Utrecht, Netherlands}
\affil[2]{University of Minnesota, Minneapolis, USA}
\affil[3]{Dartmouth College, Hanover, USA}
\affil[4]{Sharif University of Technology, Tehran, Iran}
\affil[5]{ISTA, Klosterneuburg, Austria}

\date{}
\maketitle

\begingroup
  \renewcommand{\thefootnote}{}
  \footnotetext{$\dagger$ Affiliation when work was completed. No current affiliation.}
  \footnotetext{$\star$ Equal contribution.}
\endgroup


\begin{abstract}
Spatial environmental variation can either amplify or suppress the fixation of beneficial mutants in structured populations, yet the interplay of ecological factors and spatial structure in determining which outcome occurs remains theoretically unresolved. Here, we develop a unified framework for selection on lattice graphs with environmental heterogeneity, in which mutant and resident fitness depend on the local environmental state. Across three common classes of genotype-environment interactions and a wide range of spatial arrangements of environmental states, we identify two governing principles. Genotype specificity determines the direction of the effect: heterogeneity amplifies selection when it modulates resident fitness, but suppresses selection when it modulates mutant fitness, with genotype-symmetric modulation producing weaker amplification. Spatial arrangement determines the magnitude: intermixed versus clustered environments tune the strength of amplification or suppression without reversing the direction of the effect. Together, these principles reconcile disparate theoretical results and provide predictive criteria for adaptation in heterogeneous landscapes, from microbial communities to somatic evolution and cancer.
\end{abstract}

    
\section{Introduction}
New mutations arise and spread in space and time under variable ecological conditions. 
Whether a mutation ultimately fixes or goes extinct depends not only on its intrinsic 
fitness advantage, but also on the environment it encounters and the spatial context in 
which selection operates. This interplay between ecology, space, and evolution shapes 
outcomes as diverse as the emergence of antibiotic resistance, the progression of cancer, 
and adaptation in fragmented landscapes~\cite{durrett2008probability,nagylaki1992introduction,hanski1998metapopulation,
ewens,wiens2000ecological,levins1968evolution}.

Environmental heterogeneity is a ubiquitous feature of most biological systems~\cite{kolasa1991ecological,wiens2000ecological,levins1968evolution}. Solid 
tumors contain sharp spatial gradients in oxygen and nutrient concentration, drug levels, 
and acidity~\cite{gillies2008heterogeneity,tredan2007drugpenetration,
minchinton2006drugpenetration,semenza2003hypoxia}, creating niches that favor distinct 
phenotypes. Similarly, microbial communities that grow in biofilms or on surfaces 
experience nutrient depletion zones, metabolic cross-feeding, and antibiotic gradients 
that generate complex spatial ecological mosaics~\cite{stoodley2002biofilms,
steenackers2016experimental}. These patterns of heterogeneity modulate selective 
pressures and the likelihood that beneficial or deleterious mutants establish a new 
colony. Understanding when such heterogeneity promotes or impedes the fixation of new 
variants is therefore a question of broad ecological and biomedical relevance.

Evolutionary graph models have provided a powerful framework for studying how population 
structure shapes mutant fate, showing that neighborhood topology and dispersal patterns 
can strongly modulate mutant success~\cite{nowak2006evolutionary,broom2014game,lieberman2005evolutionary,maruyama1970fixation,
maruyama1,maruyama2,traulsen2009stochastic,komarova2006spatial,hindersin2014counterintuitive,hindersin2015most,
broom2011evolutionary,broom2008analysis,monk2014martingales,houchmandzadeh2010alternative,
houchmandzadeh2011fixation,allen2015molecular,kaveh2014duality,tkadlec2019population,
pavlogiannis2017amplification,pavlogiannis2018construction,adlam2015amplifiers}. 
Examples include somatic evolution in epithelial tissues and cancer, microbial 
adaptation, and the evolution of antibiotic resistance~\cite{baym2016spatiotemporal,
vermeulen2014stem,vermeulen2013defining,souque2024petri,zhang2011acceleration,
bos2018bacterial}.

However, theoretical predictions on how environmental heterogeneity affects mutant success remain 
inconsistent. In deme-structured metapopulation models under weak selection, spatial 
variation in selection coefficients can increase fixation probability in both the weak- 
and strong-migration limits~\cite{Gavrilatzlets2002fixation,whitlock2005probability,
tachida1991fixation}. In contrast, heterogeneous fitness landscapes on complete 
graphs~\cite{kaveh2019environmental,hauser2014heterogeneity} or disordered 
lattices~\cite{manem2015modeling} tend to suppress selection. However, random fitness 
distributions in low-dimensional lattices have been shown to increase fixation 
probability~\cite{mahdipour2017genotype,farhang2017effect,farhang2019environmental}. 
Similarly, studies of spatially explicit fitness distributions show that random or 
alternating environmental states can amplify 
selection~\cite{kaveh2020moran,nemati2023counterintuitive,svoboda2023coexistence}.

These divergent outcomes highlight that population structure, environmental distribution, 
and selection regime jointly shape evolutionary outcomes in nontrivial ways. However, 
existing frameworks typically do not systematically account for the role of 
genotype-specific interactions, specifically, whether environmental variation acts 
symmetrically or asymmetrically on competing genotypes. This limits their ability to 
reconcile these results across models (with some 
exceptions:~\cite{mahdipour2017genotype,farhang2017effect}).

To address the lack of a unifying framework, we develop a model that integrates 
population structure, spatial environmental heterogeneity, and a general form of 
genotype--environment interactions within a unified selection framework. We explicitly 
distinguish between different modes of genotype specificity in how environmental 
variation affects fitness. We consider selection in graph-structured populations in which 
the fitness of competing genotypes, mutants (A) and residents (B), is shaped by the 
local environmental state under Moran birth--death dynamics. The environment is 
represented as a scalar field that assigns an environmental quality to each node 
(habitat) in the system. In the simplest case of binary environmental states, these are 
denoted as ``rich'' and ``poor'' states, indicating whether the environment increases or 
decreases fitness~\cite{kaveh2020moran}. In graph-theoretic terms, this environmental 
field corresponds to a coloring of the graph~\cite{kaveh2020moran,maciejewski2014environmental}.

The fitness of an individual at a given node consists of an inherent component and an 
environmental component (Fig.~1)~\cite{levene1953genetic,hauser2014heterogeneity,
kaveh2020moran}. The inherent component reflects the baseline fitness of a genotype in a 
uniform environment. The environmental component is a linear response to local 
environmental quality, scaled by a genotype-specific factor $\sigma_{\rm A}$ or 
$\sigma_{\rm B}$ (the heterogeneity amplitude), which determines the strength and 
direction of the environmental variation on selection.

We study three classes of genotype-environment interaction that capture the most common 
biological scenarios: genotype-symmetric environments (S1), where environmental variation 
affects both genotypes equally ($\sigma_{\rm A}=\sigma_{\rm B} = \sigma$); mutant-specific 
environments (S2), where only the mutant responds to environmental variation 
($\sigma_{\rm A}= \sigma >0$, $\sigma_{\rm B}=0$); and resident-specific environments 
(S3), where only the resident genotype is affected ($\sigma_{\rm A}=0$, $\sigma_{\rm B} 
= \sigma >0$) (Fig.~1). These scenarios encompass the principal ecological and biological 
settings in which spatial heterogeneity influences evolutionary dynamics.

To investigate how environmental heterogeneity shapes evolutionary outcomes, we vary both 
the heterogeneity amplitude and the spatial arrangement of environments on one- and 
two-dimensional lattices (Fig.~2). For each interaction scenario, S1, S2, or S3, we 
consider environmental fields spanning the full range from highly intermixed 
(checkerboard-like) to highly clustered (segregated domain). We quantify the spatial 
arrangement of environments using a single parameter, the spatial correlation index, 
which measures how strongly similar environmental states cluster in space.

For a one-dimensional cycle graph, we compute the fixation probability of a beneficial 
mutant across the three interaction scenarios and 946 environmental 
configurations that span a continuum of spatial correlations. In two dimensions, we 
complement this analysis by examining representative limiting configurations. By 
comparing heterogeneous landscapes with homogeneous baselines, we identify two general 
principles governing evolutionary outcomes. First, genotype specificity is the primary 
determinant of amplification: when environmental variation affects residents, it 
consistently amplifies mutant success, whereas variation affecting mutants suppresses it; 
genotype-symmetric heterogeneity yields a more modest amplification. Second, spatial 
arrangement acts as a secondary determinant: the degree of clustering or interleaving 
modulates the magnitude of these effects and can either strengthen or attenuate selection 
depending on the interaction scenario.

Taken together, these results show that seemingly contradictory outcomes in the 
literature can be understood within a single framework: genotype specificity determines 
the direction of selection, while spatial environmental arrangement sets its magnitude. 
This explains why heterogeneous environments can either amplify or suppress selection 
across models and spatial structures and fitness distributions, and provides criteria for 
predicting which outcome will occur. 

In the following, we define the model, derive these principles analytically for limiting 
configurations, and validate them numerically across a continuum of spatial environments. 


\section{Model} 
\subsection{Population structure and Moran dynamics}
We consider a finite population of size $N$ that resides on an undirected graph $G = (V, E)$ with a node set $V$ and an edge set $E$. Each node represents a habitat occupied by a single individual, and each edge encodes neighborhood and dispersal patterns. We focus on one-dimensional and two-dimensional undirected lattice graphs (cycle and square lattice). At any time, each node $i \in V$ hosts a mutant of type $A$ or a resident of type $B$. The state of the system is described by a binary vector $n = (n_1,\dots,n_N)$, where $n_i = 1$ if the site $i$ contains a mutant, and $n_i = 0$ otherwise.

Evolution proceeds according to a discrete-time birth-death Moran process. At each time step:
(1) an individual is chosen for reproduction with probability proportional to its fitness; and
(2) one of its neighbors is selected uniformly at random and replaced by the offspring.
This process continues until the mutant lineage either goes extinct or reaches fixation.

\subsection{Spatial fitness landscape}
Fitness depends on both genotype and local environmental state. Let $r_{\alpha}$ denote the baseline (inherent) fitness of the genotype $\alpha \in \{A,B\}$, and let $\sigma_{\alpha}$ denote the strength of its interaction with the environment. The environment is represented by a binary field $c_i \in \{+1,-1\}$, indicating whether this location is resource-rich or resource-poor. Thus, the fitness of genotype $\alpha$ at site $i$ is (Fig. 1B)
\[
f_{\alpha,i} = r_{\alpha} + \sigma_{\alpha} c_i.
\]

We assume an equal number of rich and poor nodes in all configurations so that the mean of $c_i$ is zero. This ensures that $r_{\alpha}$ represents the average fitness of genotype $\alpha$ in the landscape, while $\sigma_{\alpha}$ determines the amplitude of genotype-specific environmental variation. In our three interaction scenarios, $\sigma_{\alpha}$ is set to the value $\sigma$ whenever it is non-zero. Without loss of generality, we set $r_{\rm B}=1$ and write $r_{\rm A}=r$. The three genotype-environment interaction scenarios are:
\[
\begin{aligned}
&\text{S1 (genotype symmetric):}\quad f_{\rm A,R/P} = r \pm \sigma,\quad f_{\rm B,R/P} = 1 \pm \sigma,\\
&\text{S2 (mutant-specific):}~~~~\quad\quad f_{\rm A,R/P} = r \pm \sigma,\quad f_{\rm B,R/P} = 1,\\
&\text{S3 (resident-specific):}~~~\quad\quad f_{\rm A,R/P} = r,~\qquad\;\; f_{\rm B,R/P} = 1 \pm \sigma,
\end{aligned}
\]
where R(P) denotes rich (poor) sites, corresponding to the $+$ ($-$) sign. 

\subsection{Spatial correlation index}
To quantify the level of spatial mixing or clustering of resource-rich and resource-poor sites, we define (environmental) spatial correlation index
\begin{equation}
    M = \frac{1}{Z} \sum_{i,j} \frac{c_i c_j}{\delta(i,j)^a},
\end{equation}
where $\delta(i,j)$ is the graph distance between nodes $i$ and $j$, $a > 0$ is a decay exponent, and $Z$ is a normalization constant. We fix $a=2$ and $Z=N$ throughout. This index measures spatial correlation in the environmental field. Low values of $M$ correspond to highly intermixed environments, whereas high values
of $M$ indicate strong spatial clustering. The absolute values of $M$ depend on the size of the graph and the decay exponent $a$, but the ordering of the configurations is robust. For $a=2$, this bound would shift to approximately $-0.8$ and $+1.4$. For a further discussion of the definition and generality of $M$, see Supplementary Note 2. 

\subsection{Fixation probability}
Let $\rho_{\rm A} = \rho_{\rm A}(r,\sigma,{\bf c})$ denote the fixation probability of a single mutant of type $A$ introduced at a randomly chosen site. 
The vector ${\bf c}=(c_1,c_2,\cdots,c_N)$ represents the environmental field, $c_i$. For each environmental configuration, we compute the fixation probability of the birth-death Moran process using exact numerical solutions of the backward Kolmogorov equation in the 1D case and stochastic simulations in the 2D case. We compare these values with the homogeneous baseline ($\sigma = 0$) to quantify how much environmental heterogeneity amplifies or suppresses selection. Note that the examined graphs are isothermal, meaning that in the absence of environmental variation the fixation probability coincides with the classical well-mixed Moran result~\cite{lieberman2005evolutionary,kaveh2014duality,svoboda2024amplifiers}.


\section{Results}

\subsection{Analytical results for limiting configurations}
To establish theoretical reference points for how environmental heterogeneity modulates fixation, 
we first analyze two limiting spatial configurations: (i) a maximally mixed `checkerboard' environment 
and (ii) a fully `segregated' environment in which rich and poor regions form contiguous domains. These 
cases represent the extremes of the spatial correlation index, $M$ (Fig.~2) and provide bounds on how the 
fixation probability depends on the heterogeneity amplitude~$\sigma$ under each 
genotype-environment interaction scenario.

\textit{Checkerboard configuration.}
In a perfectly alternating landscape (checkerboard), every individual is surrounded by neighbors in 
the opposite environmental state. This symmetry allows the fixation probability to be computed analytically 
using the martingale method (Supplementary Note~3, \cite{kaveh2020moran}),
\begin{align}
\rho_{\rm A}^{\mathrm{(chk)}} 
&= \frac{\displaystyle 1 - \frac{1}{2} \left( 
\frac{f_{\rm B,R}}{f_{\rm A,R}}\cdot\frac{f_{\rm A,R} + f_{\rm B,P}}{f_{\rm A,P} + f_{\rm B,R}}
+\frac{f_{\rm B,P}}{f_{\rm A,P}}\cdot\frac{f_{\rm A,P} + f_{\rm B,R}}{f_{\rm A,R} + f_{\rm B,P}} 
\right)}
{\displaystyle 1 - \left(\frac{f_{\rm B,R} f_{\rm B,P}}{f_{\rm A,R} f_{\rm A,P}}\right)^{N/2}}\,,
\end{align}
with simplified expressions for each scenario provided in the Supplementary Note~1, Eqs.~(S2)–(S4). 
Note that $\rho_{\rm A}^{\mathrm{(chk)}}$ is identical on one-dimensional cycles and two-dimensional 
lattices of equal size, due to the environmental isothermal theorem~\cite{kaveh2020moran}. Because each 
of such configurations corresponds to a proper two-coloring of the graph, the checkerboard 
environment is also referred to as a “two-chromatic” configuration.

\textit{Segregated configuration.}
When rich and poor sites form separate domains, individuals experience spatially uniform environments within each domain. The success of a mutant depends on whether it originates in a favorable or unfavorable region. We estimate the fixation probability by treating each region as locally homogeneous and averaging the fixation probabilities across two domains.
\[
\rho_{\rm A}^{\mathrm{(seg)}} \approx 
1 - \frac{1}{2}\left(
\min\!\left(1,\frac{f_{\rm B,R}}{f_{\rm A,R}}\right)
+
\min\!\left(1,\frac{f_{\rm B,P}}{f_{\rm A,P}}\right)
\right).
\]
This approximation is valid when the product of fitness ratios satisfies
\[
\frac{f_{\rm A,R}}{f_{\rm B,R}}\cdot \frac{f_{\rm A,P}}{f_{\rm B,P}} > 1.
\]
If this product is less than unity, fixation becomes exponentially unlikely in $N$ (Supplementary Note~4). Scenario-specific expressions for S1–S3 are provided in the Supplementary Note~1, Eqs.~(S5)–(S7).

\textit{Interleaved environment.}
The interleaved configuration, defined here for one-dimensional cycle graphs, consists of a periodic arrangement in which local neighborhoods contain equal numbers of rich and poor sites. Such configurations are structurally similar to random environments because, in a fully random assignment, the expected number of rich and poor neighbors for a randomly chosen node is also equal. Analytical expressions and asymptotic behavior for this configuration are provided in Supplementary Note 5.  

Together, these limiting configurations delineate the ranges of fixation probabilities achievable for a given $(r,\sigma)$ and provide theoretical benchmarks to interpret the numerical and simulation results in the following sections. Checkerboard environments match simulations exactly, while in segregated landscapes the analytical approximation agrees qualitatively and becomes increasingly accurate for larger $N$ (Supplementary Notes~3–5).

\subsection{{\bf Genotype specificity} of the environment is the principal determinant of amplification or suppression}

Figure~\ref{fix_prob_fig} (and Figure S4) shows analytical and numerical results for the fixation probability $\rho_{\rm A}$ as a function of the heterogeneity amplitude $\sigma$ across the three interaction scenarios, in 1D cycles and 2D square lattices
with checkerboard and segregated configurations. Qualitative behavior is consistent across selection regimes ($r = 1.1, 1.5, 2$). Results for $r=1.5$ are similar to those for $r=2$ and are therefore omitted for visual clarity. As noted
in the previous subsection, in checkerboard environments $\rho_{\rm A}$ is identical for cycles and square lattices~\cite{kaveh2020moran}.

\textit{Scenario 3.} 
When the environment affects only the residents, heterogeneity amplifies selection. In all selection regimes, $\rho_{\rm A}$ increases monotonically with $\sigma$ in both checkerboard and segregated configurations. In weak selection, this amplification is substantial: for example, $\rho_{\rm A}(r=1.1, \sigma \to 1)$ is $3-4$ times larger than its value at $\sigma=0$. This trend is robust to spatial structure and environmental configurations. Similar amplification has been observed on small-$N$ complete graphs~\cite{kaveh2019environmental}. In complete graphs, the magnitude of this effect diminishes rapidly as the population size increases.

\textit{Scenario 2.}
When the environment acts only on mutants, heterogeneity suppresses selection, with a subtle exception for segregated environments.
Across both lattice structures and in both limiting configurations, $\rho_{\rm A}$ decreases as $\sigma$ becomes sufficiently large. In weak selection, strong heterogeneity can drive the fixation probability to zero. This is expected since at $\sigma = r$ the mutant fitness in one domain becomes zero, leading to $\rho_{\rm A}=0$. However, in a segregated environment, the response is non-monotonic: small heterogeneity 
initially increases $\rho_{\rm A}$ before stronger variation suppresses fixation. A similar non-monotonic behavior has been reported in spatial fitness gradients \cite{svoboda2025gradient}. This non-monotonicity and the broader similarity between Scenario~2 and Scenario~3 in the weak-selection regime, where the fixation probability can increase relative to the homogeneous model, are consistent with the diffusion-approximation  meta-population results 
\cite{Gavrilatzlets2002fixation,whitlock2005probability}, in which genotype-specificity is ambiguous (see Supplementary Note~6). Aside from this special case, the dominant trend is that mutant-specific environments reduce the probability of fixation. A similar trend is observed in complete graphs with arbitrary mutant fitness heterogeneity \cite{kaveh2019environmental}.

\textit{Scenario 1.}
When the environment affects both genotypes equally, fixation probability still increases with $\sigma$, but the effect is weaker than in resident-specific environments. Because mutants and residents experience parallel environmental variation, heterogeneity does not introduce a directional fitness bias; amplification arises only through second-order effects at the boundaries between rich and poor neighborhoods. This effect disappears under inherent neutrality ($r=1$) and in the checkerboard configuration, where symmetry eliminates boundary asymmetries \cite{kaveh2020moran}. Similarly, in the fully-segregated configuration, the curves in Fig.~3 (panels C and D, $r=1.1$) show a somewhat weak response that remains close to the homogeneous baseline.

\subsection{{\bf Spatial~environmental} arrangement modulates the magnitude of amplification or suppression}

We now ask how the spatial arrangement of environmental states, quantified by the spatial correlation index, $M$, modifies the fixation outcomes. For this part of the study we focus on the 1D cycle graph. For each value of $\sigma$, we generated 946 environmental configurations on the 1D cycle ($N=64$), spanning the full range from maximally intermixed (checkerboard, low $M$) to fully clustered (segregated domains, high $M$). The configuration details and generation schemes are provided in the Supplementary Note~2. For each configuration, $\rho_{\rm A}$ was computed for $r=1.5$, using exact numerical solutions of the backward Kolmogorov equation.

\textit{Scenario 3.}
When the environment affects only the residents, heterogeneity acts as an amplifier of selection. Across spatial structures and values of $M$, $\rho_{\rm A}$ increases with $\sigma$, but the magnitude of amplification strongly depends on $M$. Figure~\ref{FP_vs_M_fig}C shows the strongest amplification in highly intermixed environments (low-$M$) and the weakest amplification in clustered landscapes (high-$M$).

\textit{Scenario 2.}
Here, the trend reverses: fixation probability increases with $M$ (Figure~\ref{FP_vs_M_fig}B). Poor mixing or segregation reduces the exposure of a mutant to unfavorable sites and increases $\rho_{\rm A}$, whereas highly intermixed environments impose frequent environmental switches and suppress fixation. Thus, spatial clustering counteracts the suppressive effect of mutant-specific heterogeneity.

\textit{Scenario 1.}
When both genotypes respond equally to the environment, the dependence on $M$ is non-monotonic (Figure~\ref{FP_vs_M_fig}A). From a checkerboard configuration, partial clustering decreases $\rho_{\rm A}$, but further clustering increases it again. Checkerboard and segregated landscapes therefore yield similar fixation probabilities for a given $\sigma$, with a minimum at intermediate values of the spatial correlation index $M$. This minimum is established analytically for the ``interleaved" configuration in the Supplementary Note~5.

Supplementary Figures~S1–S2 further illustrate these trends. Figure~S1 shows representative configurations across $M$, while Figure~S2 provides an approximate contour map of $\rho_{\rm A}$ in the $(M,\sigma)$ plane. Figure S3 shows the location of approximate extrema for the fixation probability. 

Together, these results provide a unified framework that includes most of the reported results in the literature. Figure~5, and Table S1, summarize the amplification and suppression regimes and indicate where previous studies in the literature fall within this landscape. (See also Supplementary Note 6 for a detailed overview of the previous studies.) Studies reporting amplification align with regions of resident-specificity, while those reporting suppression correspond to mutant-specificity. 

\subsection{Remarks on graph structure and connectivity}

The results above were obtained on low-connectivity lattices (cycles and square lattices), yet the qualitative dependence on genotype specificity is not restricted to these structures. On complete graphs with arbitrary location-dependent fitness, mutant-specific heterogeneity suppresses selection, whereas resident-specific heterogeneity produces a weak amplification that vanishes as population size increases \cite{kaveh2019environmental}. The role of the spatial correlation index, however, does not generalize to high-connectivity graphs. As degree approaches the complete-graph limit, spatial correlations in the environment lose relevance, and the distinction between intermixed and clustered configurations becomes immaterial. In this regime, genotype specificity remains predictive, but spatial arrangement no longer modulates fixation probability in a meaningful way. These results are summarized in Supplementary Table~S2.


\section{Discussion} 

Environmental heterogeneity is a pervasive feature of biological systems, yet its influence on evolutionary dynamics has remained conceptually fragmented. Classical approaches in population genetics typically treat environmental variation as fluctuations in the effective selection coefficient, and therefore obscure distinctions between genotype-specific and genotype-symmetric effects~\cite{Gavrilatzlets2002fixation,whitlock2005probability}. Spatial graph models, on the other hand, have produced a wide range of seemingly contradictory outcomes: heterogeneity has been reported to amplify~\cite{kaveh2020moran,mahdipour2017genotype,nemati2023counterintuitive,svoboda2023coexistence} or suppress~\cite{hauser2014heterogeneity,kaveh2019environmental,manem2015modeling} selection depending on the assumptions. Our work provides a unified framework that reconciles these disparate results by explicitly incorporating genotype-environment interactions and spatial environmental arrangement into Moran dynamics on lattice graphs. Figure 5 illustrates how our model encompasses a wide range of results from the literature.

A central conclusion of our study is that the direction and magnitude of the effect of heterogeneity on the probability of fixation are primarily governed by genotype specificity: which genotype fitness is most strongly modulated by the environment. This simple principle explains the major discrepancies across earlier models. When heterogeneity disproportionately penalizes residents (resident-specific environments), selection is consistently amplified: beneficial mutants experience a more favorable competitive landscape, and fixation probability increases with heterogeneity. When heterogeneity acts on the mutant (mutant-specific environments), selection is suppressed, and fixation probability decreases, often sharply, as $\sigma$ increases. Symmetric heterogeneity, which affects both genotypes equally, produces modest amplification. These three qualitative regimes recapitulate and unify the observations reported in models related to drug gradients, nutrient variability, and landscape heterogeneity~\cite{hauser2014heterogeneity,kaveh2019environmental,farhang2017effect}.

A second key determinant of fixation is the spatial arrangement of environmental states. By sampling hundreds of configurations across the full spectrum of the environmental landscape, we show that mixing modulates but does not override the effect of genotype specificity. Resident-specific heterogeneity yields the highest fixation probabilities in highly mixed environments, while mutant-specific heterogeneity exhibits the opposite trend. In symmetric environments, fixation probability responds non-monotonically to mixing, with minima near randomly mixed configurations. These effects are captured analytically in the three limiting cases—checkerboard, segregated, and interleaved landscapes—which serve as natural benchmarks. The checkerboard case reflects a fully mixed environment where fixation depends on an effective geometric mean of fitness, reminiscent of Gillespie’s criterion for fluctuating selection~\cite{gillespie1974natural,gillespie1977natural}. Segregated landscapes represent the opposite extreme, where fixation approximates an average of two homogeneous environments. The interleaved configuration mimics a random distribution, with the same spatial correlation index. The agreement between these limiting predictions and simulation results highlights that the geometry of environmental mixing is a critical but previously underappreciated factor shaping evolutionary outcomes.

Biologically, these findings clarify when and why environmental heterogeneity accelerates or impedes selection. In tumors, drug penetration and oxygen gradients typically penalize drug-sensitive residents more than resistant mutants, creating resident-specific heterogeneity that increases the likelihood of resistant clone establishment. In microbial communities, nutrient gradients can favor strains capable of exploiting specific metabolites or spatial niches, generating mutant- or resident-specific scenarios depending on the metabolic architecture. In both systems, our results show that the spatial arrangement of these niches—whether sharply segregated or finely interleaved—can profoundly alter fixation probabilities even when the overall amount of resource is unchanged.

In general, our framework shows that environmental heterogeneity does not inherently amplify or suppress selection; rather, its effect emerges from the interplay between two major factors: genotype specificity and spatial mixing. This resolves apparent contradictions in the literature and provides a general predictive principle applicable across ecological and biomedical contexts. Future extensions may incorporate temporal fluctuations, evolving environmental landscapes, or higher-dimensional tissues and mosaics. Together, these developments promise a more complete understanding of adaptation in complex environments where spatial structure and ecological variation are essential components of evolutionary dynamics.

    
\section*{Acknowledgment}
J.S. and K.C. were supported by the European Research Council (ERC) CoG 863818 (ForM-SMArt) and Austrian Science Fund (FWF) 10.55776/COE12.
The contribution of Hossein Nemati was made before joining his current institution.
    
\bibliographystyle{unsrtnat}
\bibliography{references}
    

\newpage

\begin{figure}[h]
\begin{center}
\captionsetup{justification=raggedright,singlelinecheck=false}
\includegraphics[width=1\textwidth]{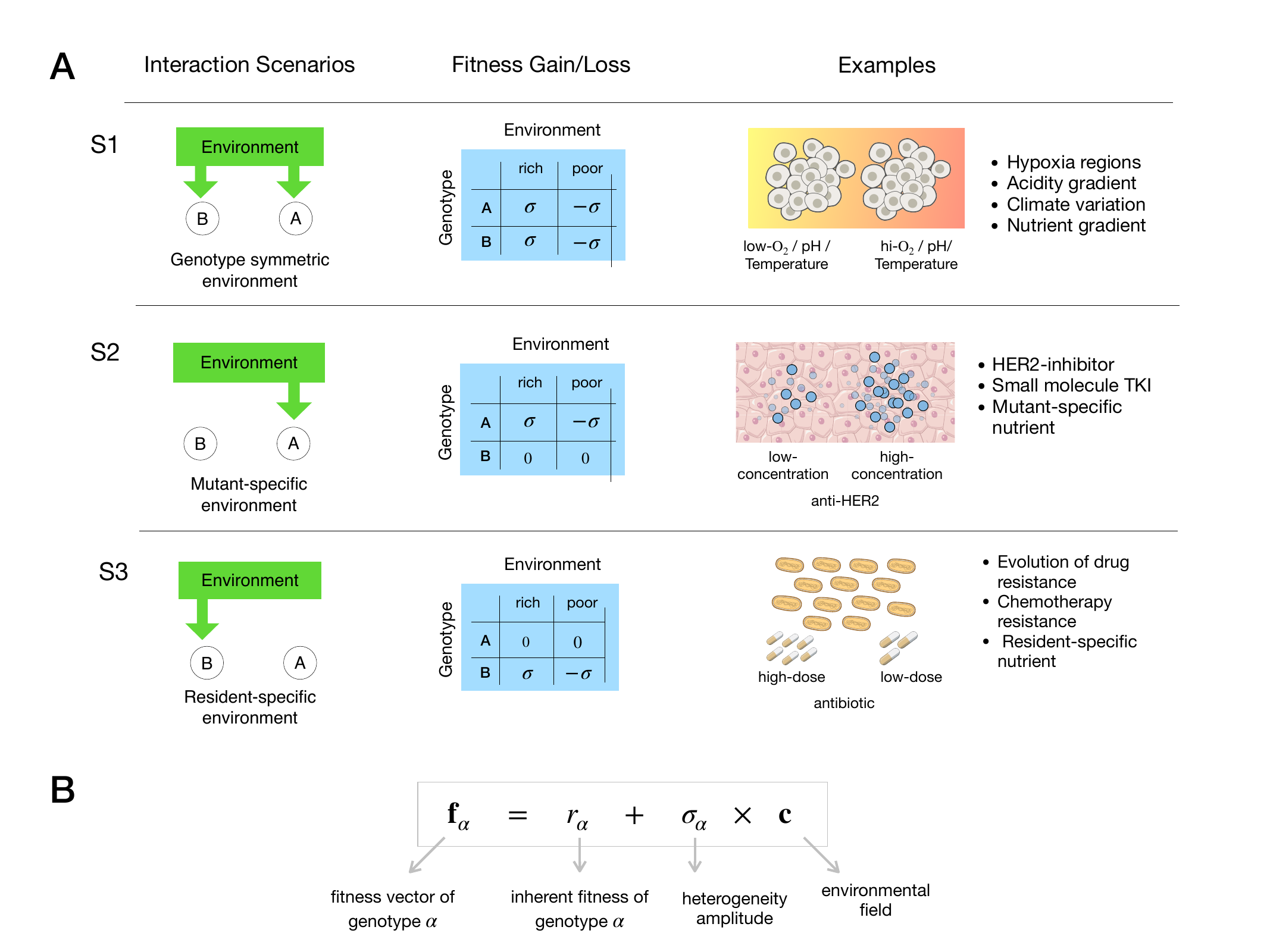}
\caption{\textbf{Genotype-environment interaction framework and the three environmental scenarios.}
(\textbf{A})~Local environmental states are rich ($c_i=+1$) or poor ($c_i=-1$), producing fitness gains or losses that depend on the genotype-environment interaction. Three scenarios reflect which genotype is affected by 
environmental variation: (S1)~genotype-symmetric environments, where both genotypes respond equally; 
(S2)~mutant-specific environments, where only mutants respond to environmental variation; and 
(S3)~resident-specific environments, where only residents respond to environmental quality. Illustrative 
biological contexts are shown for each case.
(\textbf{B})~Local fitness is decomposed into an inherent component and an environment-dependent contribution,
$f_{\alpha,i}=r_\alpha+\sigma_\alpha c_i$, where $r_\alpha$ is the inherent fitness of a genotype $\alpha$, 
$\sigma_\alpha$ is its heterogeneity amplitude, and $c_i$ is the local environmental state.}
\label{fig1}
\end{center}
\end{figure}

\begin{figure}
\begin{center}
\captionsetup{justification=raggedright,singlelinecheck=false}
\includegraphics[width=.9\textwidth]{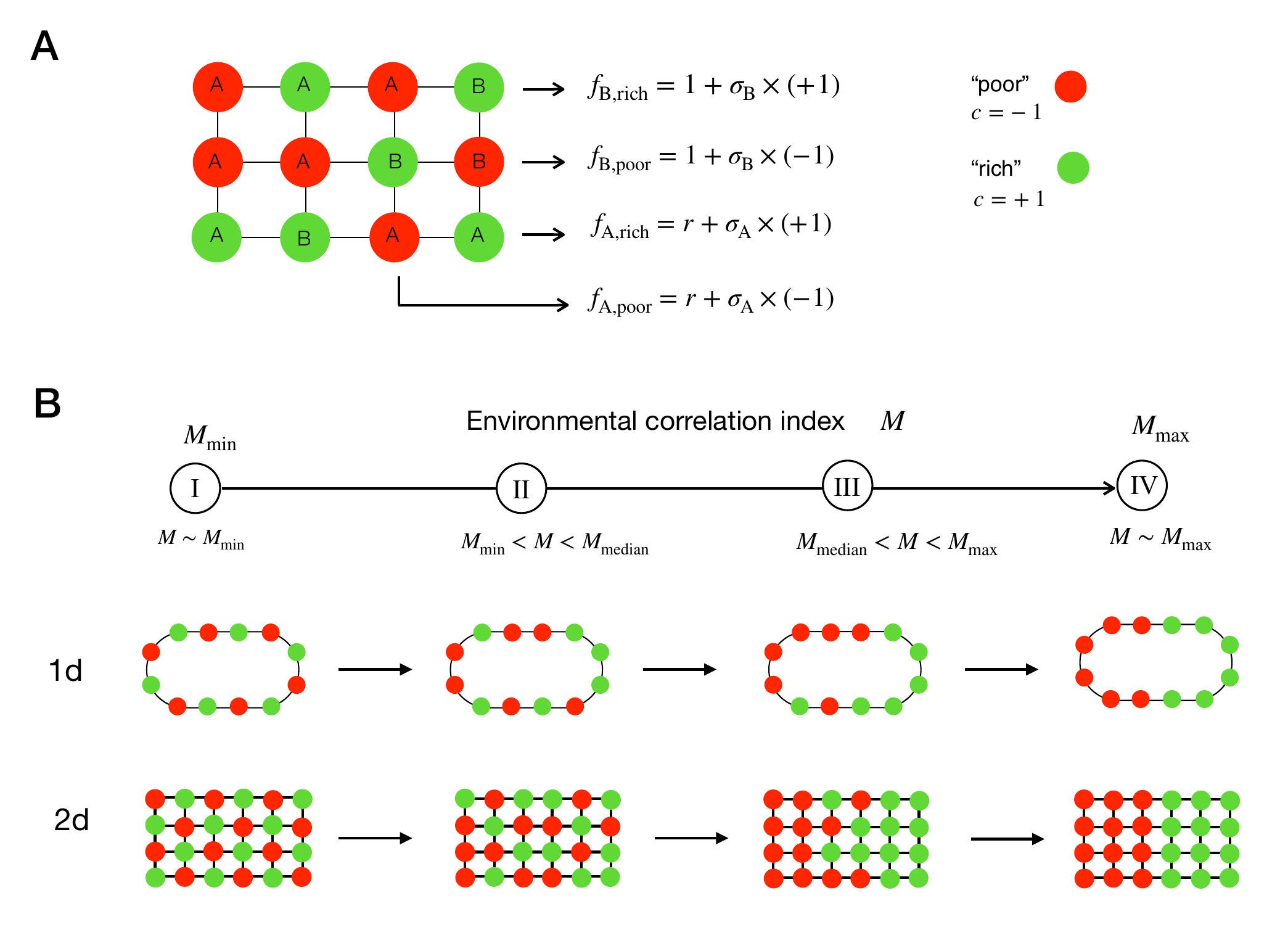}
\caption{\textbf{Spatial environmental configurations and fitness assignment.}
(\textbf{A}) Local fitness depends on the environmental state at each site: $f_{\alpha,i}=r_\alpha+\sigma_\alpha c_i$,
where $c_i\in\{+1,-1\}$ denotes rich or poor sites. Red sites correspond to $c_i=-1$ (poor) and green sites to $c_i=+1$ (rich).
The three interaction scenarios in Fig.~1 arise from different choices of $(\sigma_A,\sigma_B)$.
(\textbf{B}) Representative environmental landscapes spanning the full range of the spatial correlation index $M$, from checkerboard configurations (maximally intermixed, low-$M$) through intermediate configurations to segregated domains (highly clustered, high-$M$). Examples are shown for both 1D cycles and 2D lattices.}
\label{fig2}
\end{center}
\end{figure}

\begin{figure}
\begin{center}
\includegraphics[width=1\textwidth]{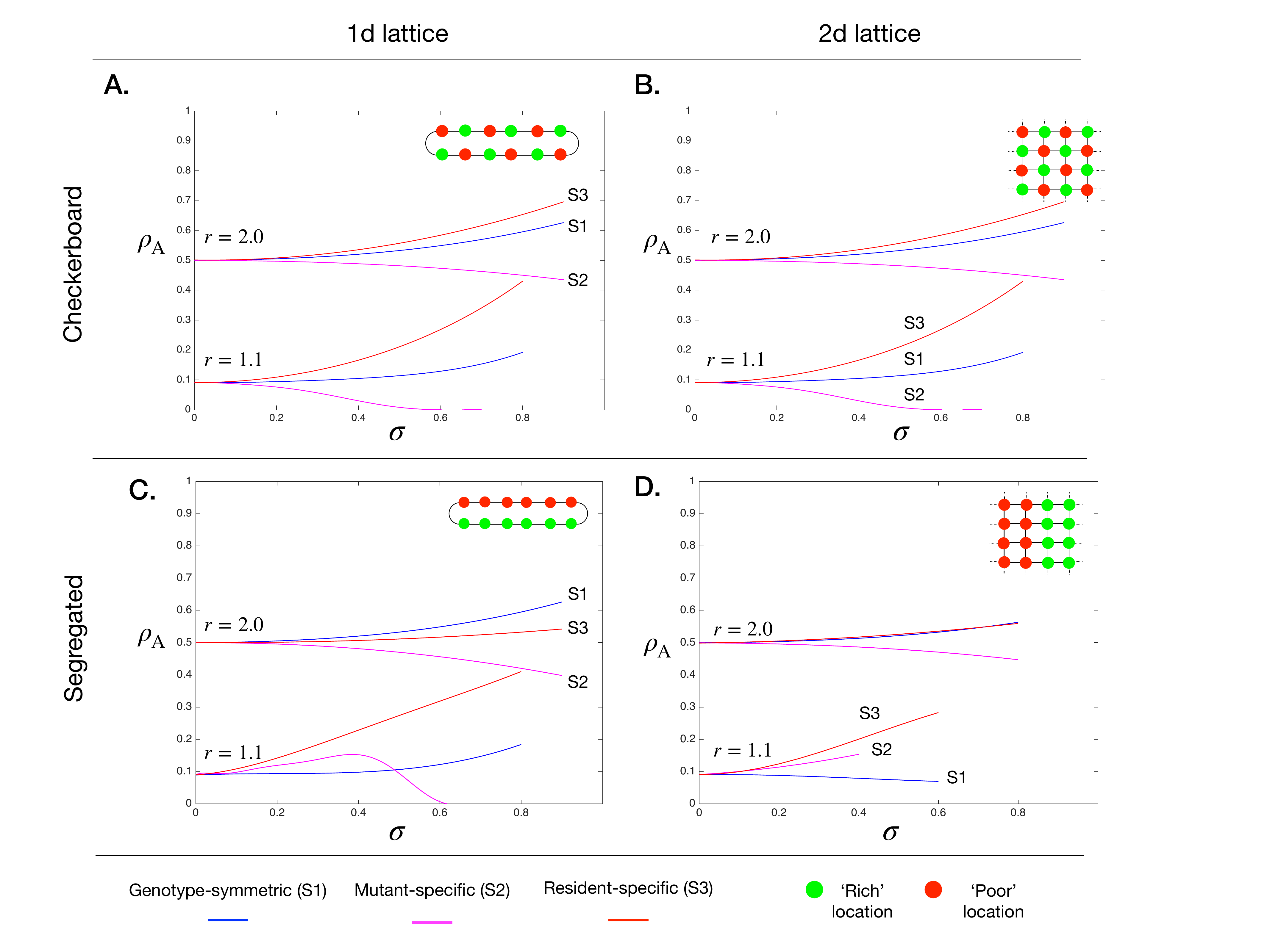}
\caption{\textbf{Fixation probability as a function of heterogeneity amplitude.}
Fixation probability $\rho_{\rm A}$ as a function of $\sigma$ for the three
genotype-environment interaction scenarios (S1–S3). Panels correspond to:
(\textbf{A}) 1D cycle, checkerboard (maximally intermixed, low $M$);
(\textbf{B}) 2D lattice, checkerboard (maximally intermixed, low $M$);
(\textbf{C}) 1D cycle, segregated (highly clustered, high $M$);
(\textbf{D}) 2D lattice, segregated (highly clustered, high $M$).
Each panel shows results for two selection strengths ($r=1.1$ and $r=2.0$, see Fig.~S4 for $r=1.5$).
Solid curves denote numerical or simulation results; in checkerboard environments, they coincide with the analytical prediction (see Results: Analytical results in limiting configurations and Supplementary Note 3). Population size $N=64$.
}
\label{fix_prob_fig}
\end{center}
\end{figure} 

\begin{figure}
\begin{center}
\captionsetup{justification=raggedright,singlelinecheck=false}
\includegraphics[width=0.8\textwidth]{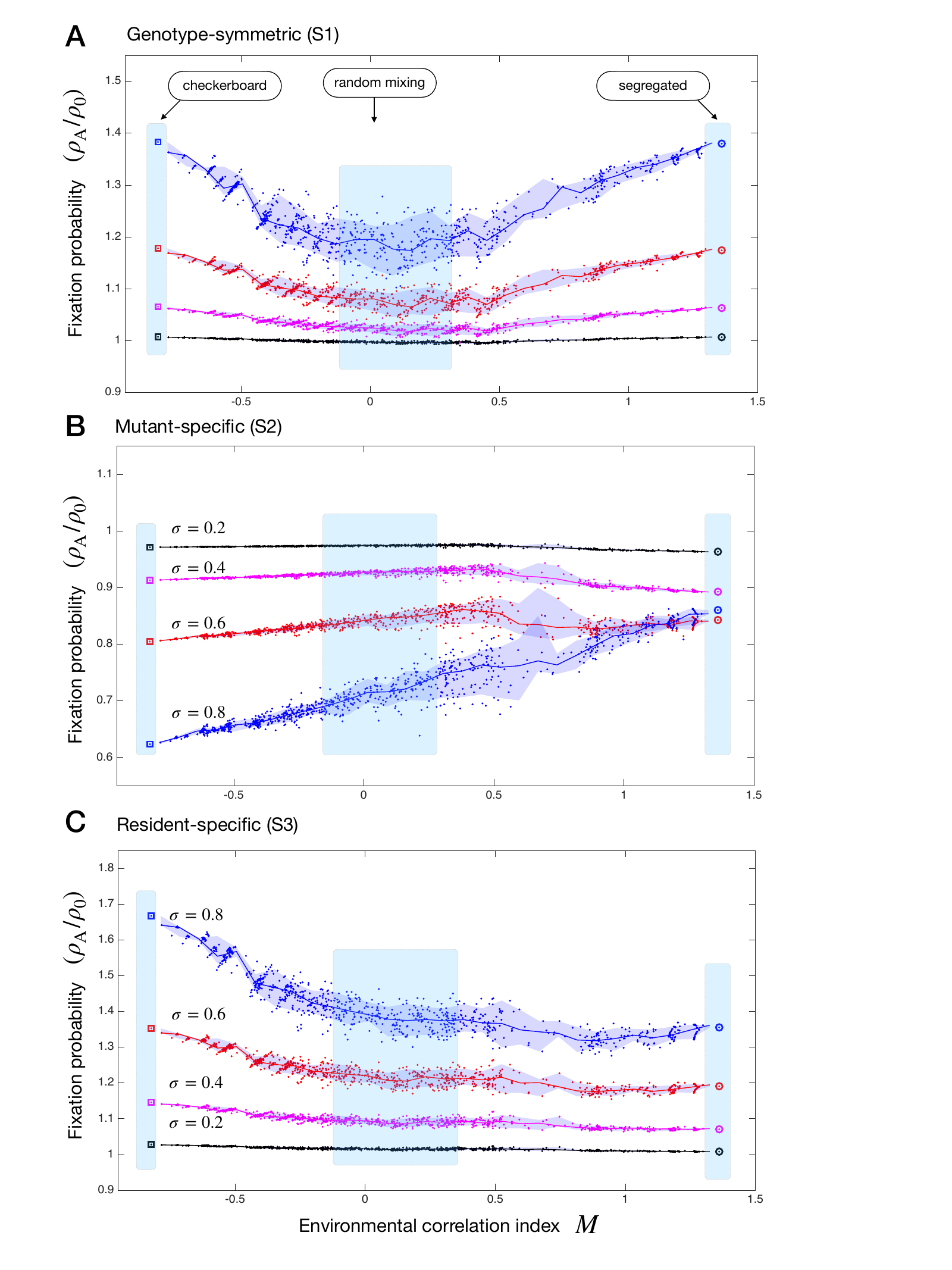}
\caption{\textbf{Dependence of fixation probability on spatial arrangement of environments.} 
Fixation probability $\rho_{\rm A}$ versus spatial correlation index, $M$, for representative 
heterogeneity levels $\sigma$ under the three genotype-environment interaction 
scenarios (S1–S3). 
Each point corresponds to one of 946 distinct environmental configurations on a 
1D cycle ($N=64$) with baseline fitness $r=1.5$. 
Checkerboard and segregated landscapes occupy the extremes of the $M$ axis, while values near the median value of $M$ correspond to uncorrelated or randomly mixed environmental arrangements. Colored envelopes indicate the 10th–90th percentile range of fixation probabilities across environmental configurations at fixed $\sigma$.(See also Figure S1-Figure S3 in the Supplementary Note 7.)}
\label{FP_vs_M_fig}\label{fig4} 
\end{center}
\end{figure}

\begin{figure}
\begin{center}
\captionsetup{justification=raggedright,singlelinecheck=false}
\includegraphics[width=.8\textwidth]{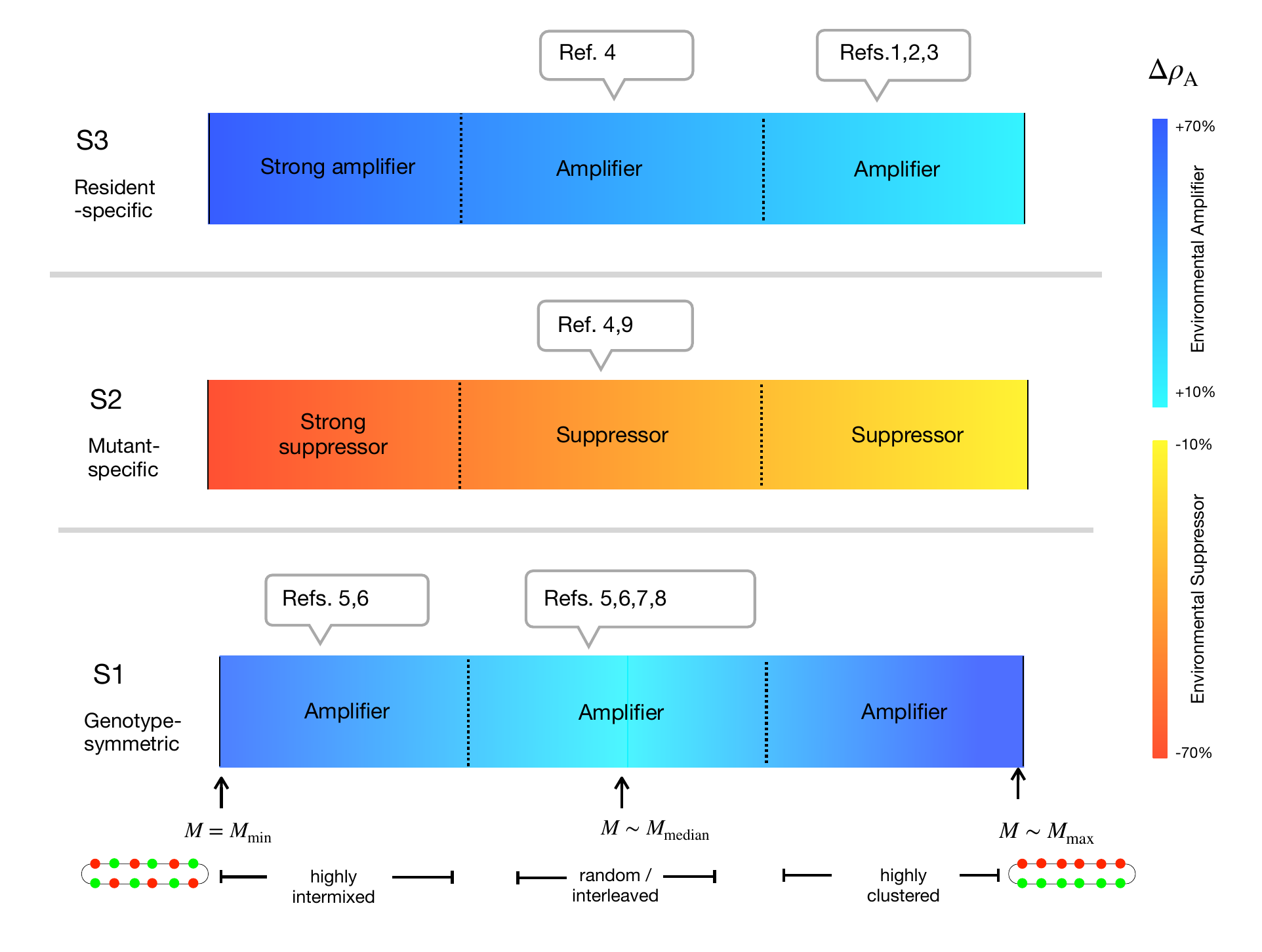}
\caption{\textbf{Unified classification of amplification and suppression by environmental heterogeneity.}
Schematic summary of how environmental heterogeneity affects the fixation probability of beneficial mutants across genotype-environment interaction scenarios and spatial environmental arrangements.
Colors indicate the relative change in fixation probability $\Delta\rho_{\rm A}$ compared to the corresponding homogeneous environment, ranging from strong suppression (blue) to strong amplification (red).
Rows correspond to the three genotype-specific interaction scenarios (S1–S3), while columns legend represent limiting environmental configurations: checkerboard (maximally intermixed, low $M$), interleaved or random (intermediate $M$), and segregated (highly clustered, high $M$).
Reference numbers indicate representative studies from the literature mapped onto each regime (see Supplementary Table~S1 for references and Supplementary Note~6 for review of the literature).
The figure illustrates that genotype specificity determines the direction of heterogeneity’s effect (amplification versus suppression), whereas spatial arrangement primarily modulates its magnitude.
}
\label{table1}
\end{center}
\end{figure}

\end{document}


\title{Supplementary Information for\\[0.3em]
\large \textit{Genotype specificity and spatial arrangement govern the direction and magnitude of selection in variable environments}}
\author{Hossein Nemati, Kamran Kaveh, Jakub Svoboda, Krishnendu Chatterjee \and Mohammad Reza Ejtehadi}

\date{}
\maketitle

\begin{center}
This PDF contains Supplementary Notes S1--S7, including Supplementary Figures and Tables.
\end{center}

\vspace{1em}

\tableofcontents
\begin{itemize}
\item[] Supplementary Note 1. Summary of analytical results.
\item[] Supplementary Note 2. Spatial correlation index $M$
\item[] Supplementary Note 3. Checkerboard configuration: Martingale Method 
\item[] Supplementary Note 4. Segregated configuration
\item[] Supplementary Note 5. Interleaved configuration
\item[] Supplementary Note 6. Overview of previous findings in the literature  
\item[] Supplementary Note 7. Supplementary Figures and Tables 
\end{itemize}
\newpage

\section*{Supplementary Note 1: Summary of Analytical Results}

In this section, we summarize the analytical expressions for the fixation 
probability $\rho_{\rm A}$ under the three limiting environmental configurations 
examined in the main text: (i) the maximally mixed \emph{checkerboard},  
 (ii) fully \emph{segregated} configurations, and (iii) \emph{interleaved} (“2–2”) environment. 
The derivations appear in Sections~S3–S5. The checkerboard configuration results are applicable to 1D cycle and 2D square lattice graph while segregated and interleaved analytical results are limited to the cycle graph. 

\subsection*{S1.1 Checkerboard environment}

For a checkerboard configuration in which rich and poor sites 
strictly alternate, the fixation probability of a single mutant is given by 
the following closed-form expression:


\begin{align}
\rho_{\rm A}^{\mathrm{(chk)}} 
&= \frac{\displaystyle 1 - \frac{1}{2} \left( 
\frac{f_{\rm B,R}}{f_{\rm A,R}}\cdot\frac{f_{\rm A,R} + f_{\rm B,P}}{f_{\rm A,P} + f_{\rm B,R}}
+\frac{f_{\rm B,P}}{f_{\rm A,P}}\cdot\frac{f_{\rm A,P} + f_{\rm B,R}}{f_{\rm A,R} + f_{\rm B,P}} 
\right)}
{\displaystyle 1 - \left(\frac{f_{\rm B,R} f_{\rm B,P}}{f_{\rm A,R} f_{\rm A,P}}\right)^{N/2}}\,,
\end{align}

\noindent where $f_{\rm A/B, R/P}$ denote the fitness of mutant (A) and resident (B) at rich (R) and poor (P) locations. The scenario-specific closed forms are:

\begin{align}
\rho_{\rm A} &=  
\frac{1 - \Big( \frac{r-\sigma^2}{r^2 - \sigma^2}\Big)}{1 - \Big(\frac{1 - \sigma^2}{r^2-\sigma^2}\Big)^{N/2}}
&& \text{(Scenario 1: genotype-symmetric)},\\[0.5em]
\rho_{\rm A} &= 
\frac{1 - \frac12\!\left( \frac{r+1+\sigma}{(r+\sigma)(r+1-\sigma)} + 
\frac{r+1-\sigma}{(r-\sigma)(r+1+\sigma)} \right)}
{1 - \Big(\frac{1}{r^2-\sigma^2}\Big)^{N/2}}
&& \text{(Scenario 2: mutant-specific)},\\[0.5em]
\rho_{\rm A} &= 
\frac{1 - \frac12\!\left( 
\frac{1-\sigma}{r} \frac{r+\sigma+1}{r-\sigma+1}
+
\frac{1+\sigma}{r} \frac{r-\sigma+1}{r+\sigma+1}
\right)}
{1 - \Big(\frac{1-\sigma^2}{r^2}\Big)^{N/2}}
&& \text{(Scenario 3: resident-specific)}.
\end{align}

These expressions follow from a martingale construction based on the 
generating-function method (Section~S3). 

\subsection*{S1.2 Segregated configuration}

When rich and poor regions occupy contiguous blocks, fixation can be 
approximated as a combination of two homogeneous environments:
\[
\rho_{\rm A} \approx
1 - \frac12\!\left(
\min\!\left(1,\frac{f_{\rm B,R}}{f_{\rm A,R}}\right)
+
\min\!\left(1,\frac{f_{\rm B,P}}{f_{\rm A,P}}\right)
\right),
\]
valid when
\[
\frac{f_{\rm A,R}}{f_{\rm B,R}}
\cdot
\frac{f_{\rm A,P}}{f_{\rm B,P}} > 1.
\]
If this product is less than one, the fixation probability decays exponentially 
with $N$,
\(
\rho_{\rm A} \approx 2^{-\Omega(N)}.
\)

Scenario-specific approximations are:
\begin{align}
\rho_{\rm A} &\approx 1 - \frac12\!\left(\frac{1+\sigma}{r+\sigma} + \frac{1-\sigma}{r-\sigma}\right)
&& \text{(Scenario 1)},\\
\rho_{\rm A} &\approx 1 - \frac12\!\left(\frac{1}{r+\sigma} + \max\!\left(1, \frac{1}{r-\sigma}\right)\right)
&& \text{(Scenario 2)},\\
\rho_{\rm A} &\approx 1 - \frac12\!\left(\max\!\left(1,\frac{1+\sigma}{r}\right) + \frac{1-\sigma}{r}\right)
&& \text{(Scenario 3)}.
\end{align}

See Section~S4 for the full proof.

\subsection*{S1.3 Interleaved (“2–2”) environments}

For the periodic pattern of two rich sites followed by two poor sites 
(the “2–2” interleaved environment), the probability that a newly arising 
mutant dies before reproducing (Scenario~1) is,
\[
\mathbb{P}(\text{death before first reproduction})
=
\frac{r+1}{(r+1)^2 - \sigma^2}.
\]
This yields an upper bound on the fixation probability. The derivation 
is given in Section~S5.

\bigskip
\noindent
The following sections provide detailed technical derivations.

\section*{Supplementary Note 2: Spatial correlation index $M$}

In the main text, spatial mixing of environmental states is quantified by the
spatial correlation index
\[
M=\frac{1}{Z}\sum_{i,j}\frac{c_i c_j}{\delta(i,j)^a},
\]
where $c_i\in\{+1,-1\}$ denotes the binary environmental state at site $i$,
$\delta(i,j)$ is the graph distance between sites $i$ and $j$,
$a>0$ is a decay exponent, and $Z$ is a normalization constant.
Throughout this work we fix $a=2$ and use $Z=N$.

The decay exponent $a$ controls how strongly local versus long-range correlations
contribute to $M$.
For very large $a$, the index becomes dominated by nearest-neighbor correlations,
leading to many distinct configurations sharing similar values of $M$.
Using a finite value of $a$ reduces this degeneracy by incorporating correlations
over multiple length scales, at the cost of shifting the numerical range of $M$.
Importantly, for all values of $a$, random environmental configurations correspond
to intermediate values of $M$ between highly intermixed and strongly clustered
landscapes.

The index $M$ quantifies the degree of spatial correlation in the environmental
field: lower values correspond to highly intermixed arrangements, whereas higher
values indicate stronger spatial clustering of similar environmental states.
Throughout the manuscript, $M$ is used as a \emph{relative} measure of spatial
mixing rather than as a normalized quantity bounded to a fixed interval.

We verified that the qualitative results reported in the main text are robust to
the choice of spatial-mixing metric. In particular, using alternative measures of
spatial structure that we explicitly tested, nearest-neighbor Pearson correlation
(which corresponds to the large-$a$ limit of $M$) and cluster-size entropy, does
not change whether environmental heterogeneity amplifies or suppresses selection.
Although these measures differ in how they rank intermediate configurations and
rescale the spatial-mixing axis, they consistently distinguish strongly clustered
from highly intermixed environments and leave the amplification–suppression
correlation with the inter-mixing index unchanged. For this reason, $M$ serves as a minimal and sufficient
descriptor of spatial mixing for the purposes of this study.

Environmental configurations were generated by iteratively swapping rich and poor
sites, starting from either a checkerboard or a segregated configuration. This
procedure produces a broad ensemble of landscapes spanning a wide range of spatial
correlation values and allows continuous interpolation between the two limiting
cases.

Configurations with similar values of $M$ are not unique and may differ
substantially in local structure. This non-uniqueness accounts for the increased
dispersion of fixation probabilities observed for intermediate values of $M$ in
Fig.~4 of the main text.

For reference, with the parameters used in this study ($a=2$, $N=64$, $Z=N$), the
spatial correlation index takes approximate values $M\approx -0.82$ for the
checkerboard configuration, $M\approx 0.21$ for the interleaved configuration,
and $M\approx 1.36$ for the segregated configuration. These numerical values
depend on system size and normalization and are reported solely to anchor the
limiting configurations.

\newpage

\section*{Supplementary Note 3: Checkerboard Configuration: Martingale Method}\label{sec:two_chrom}

In this section, we derive exact analytical results for the checkerboard configuration. A general derivation for arbitrary bipartite graphs is given in \cite{kaveh2020moran}. 

We consider a Moran birth–death process on a graph with location-dependent fitness, following the formalism of \cite{kaveh2020moran}. Transition rates $W_i^\pm$ are defined by fitness-weighted reproduction and local dispersal.


The transition probabilities $W^{\pm}_{i}$, for gaining and losing a mutant at location $i$ are 

\begin{align}
W^{+}_{i}({\bf n}) &= {\rm Prob}(n_{1},\cdots, n_{i},\cdots, n_{N} \to n_{1},\cdots,n_{i}+1,\cdots,n_{N})\nonumber\\
&=  {\displaystyle \sum_{j}\frac{f_{{\rm A},j}n_{j}}{\sum_{k}\Big(f_{{\rm A},k}n_{k} + f_{{\rm B},k}(1-n_{k})\Big)}\times w_{ji}\big(1 - n_{i}\big)}\nonumber\\
W^{-}_{i}({\bf n}) &=  {\rm Prob}(n_{1},\cdots,n_{i},\cdots,n_{N} \to n_{1},\cdots,n_{i}-1,\cdots,n_{N})\nonumber\\
&= {\displaystyle \sum_{j}\frac{f_{{\rm B},j}(1-n_{j})}{\sum_{k}\Big(f_{{\rm A},k}n_{k} + f_{{\rm B},k}(1-n_{k})\Big)}\times w_{ji} n_{i}}
\label{transitionW}
\end{align}

Here, $w_{ji}$ denotes the probability that an offspring produced at node $j$ replaces the individual at node $i$. ${\bf n} = (n_1,\cdots n_N)$ is the vector state of the populations, and $f_{\rm A,i} (f_{\rm B,i})$ are fitness of type A (B) at location $i$ respectively.  

Defining the generating function $F(\boldsymbol{\zeta};t)$ as

\begin{align}
F(\boldsymbol{\zeta};t) = \sum_{\mathbf{n}} \zeta_{1}^{n_{1}}\cdots\zeta_{N}^{n_{N}}p(n_{1},\cdots,n_{N};t)
\end{align}

Where $p(n_{1},\cdots,n_{N};t)$ is the probability to be in the state ${\bf n} = (n_1,\cdots, n_N)$ at time $t$. Using the master equation for the Moran process, defined by the transition probabilities Eqs. \ref{transitionW}, we can obtain an equation for the generating function $F(\boldsymbol{\zeta};t)$ \cite{houchmandzadeh2010alternative,houchmandzadeh2011fixation,kaveh2020moran}

\begin{align}
\frac{\partial F(\boldsymbol{\zeta};t)}{\partial t} &= \textstyle \Big\{\sum_{ij}\Big(\big(\zeta_{j}-1\big)w_{ji}f_{{\rm A},i} + \big(\zeta^{-1}_{i}-1\big)w_{ji}f_{{\rm B},j}\Big) \hat{n}_{i}\nonumber\\
&\textstyle - \sum_{ij}\Big((\zeta_{i}-1)f_{{\rm A},j}w_{ji} + (\zeta_{i}^{-1}-1)f_{{\rm B},j}w_{ji}\Big)\hat{n}_{i}\hat{n}_{j}\Big\}\big(\hat{N}_{r}\big)^{-1}F(\boldsymbol{\zeta};t)
\label{Feq2}
\end{align}

Here, $\hat{n}_{i}=\zeta_{i}\partial/\partial\zeta_{i}$ is the mutant number operator at location $i$ and $\hat{N}_{r}$ is the total fitness operator,
$\hat{N}_{r} =  \sum_{k} f_{{\rm A},k}\Big(\zeta_{k}\frac{\partial}{\partial \zeta_{k}}\Big) + f_{{\rm B},k}\Big(1 - \zeta_{k}\frac{\partial}{\partial \zeta_{k}}\Big)$. If the value of $F(\boldsymbol{\zeta};t)$ is independent of time for some given $\boldsymbol{\zeta} = \boldsymbol{\zeta^{\star}}$ it defines a martingale. The value of fixation probability is obtained from the formula 

\begin{align}
\textstyle
\rho_{A}({\bf f}_{\rm A},{\bf f}_{\rm B}) = {\displaystyle \frac{1 - (1/N){\textstyle \sum_{i}}\zeta^{\star}_{i}}{1 - \prod_{k}\zeta^{\star}_{k}}}
\label{fix_mart}
\end{align}

where ${\bf f}_{\rm A} = (f_{\rm A,1}, \cdots, f_{\rm A,N}), {\bf f}_{\rm B} = (f_{\rm B,1}, \cdots, f_{\rm B,N})$, and $f_{\rm A,i} \in \{ f_{\rm A,R}, f_{\rm A,P} \}, f_{\rm B,i} \in \{ f_{\rm B,R}, f_{\rm B,P} \}$ where R and P denote rich and poor. 

The time independence requires that in Eq.~\ref{Feq2} coefficients in front of operators $\hat{n}_{i}$ and the coefficients in front of $\hat{n}_{i}\hat{n}_{j}$ terms are zero for a set of $\zeta^{\star}$ values, i.e.

\begin{align}
\textstyle
0 &=\textstyle \sum_{j}\big(\left(\zeta^{\star}_{j}-1\right)w_{ji}f_{{\rm A},i}\big) + \left(\zeta^{\star -1}_{i}-1\right) \big(\sum_{j}w_{ji}f_{{\rm B},j}\big)
\end{align}

For a bipartite graph, due to symmetry, there are only two values for $\zeta$, that is, $ \zeta_i \in \{ \zeta_1, \zeta_2\}$. the solutions for $\zeta_1$ and $\zeta_2$ are straightforward and are reported in \cite{kaveh2020moran}. Using these $\zeta^\star$ values and substituting into Eq.~\eqref{fix_mart}, we obtain a closed-form expression for the fixation probability in the checkerboard configuration for genotype–environment interaction scenarios S1–S3:


\begin{align}
\rho_{\rm A}^{\mathrm{(chk)}} 
&= \frac{\displaystyle 1 - \frac{1}{2} \left( 
\frac{f_{\rm B,R}}{f_{\rm A,R}}\cdot\frac{f_{\rm A,R} + f_{\rm B,P}}{f_{\rm A,P} + f_{\rm B,R}}
+\frac{f_{\rm B,P}}{f_{\rm A,P}}\cdot\frac{f_{\rm A,P} + f_{\rm B,R}}{f_{\rm A,R} + f_{\rm B,P}} 
\right)}
{\displaystyle 1 - \left(\frac{f_{\rm B,R} f_{\rm B,P}}{f_{\rm A,R} f_{\rm A,P}}\right)^{N/2}}\,,
\end{align}

\noindent Upon substitution for $f_{\rm A,R/P}, f_{\rm B,R/P}$ in terms of $r$ and $\sigma$, we recover the expressions summarized in Supplementary Note 1.

\newpage

\section*{Supplementary Note 4: Segregated Configuration}\label{sec:app_segregated}

In the segregated configuration on a one-dimensional cycle, we bound the fixation probability of a single invading mutant.

To approximate the fixation probability, we simplify the dynamics in three steps. 
First, we neglect mutants that arise sufficiently close to the boundary between environments, so that the initial spread occurs effectively within a homogeneous region. Second, we compute the probability that the mutant population expands to occupy half of the heterogeneous environment. Finally, we compute the probability that the mutants successfully invade the remaining half of the graph.

To compute the fixation probability of the mutants, We use equation (6.13) from~\cite{nowak2006evolutionary}, which gives the fixation probability for a one-dimensional birth–death Markov chain. Given a one-dimensional Markov chain $v_1, v_2, \dots, v_N$, the probability that starting at $v_1$ the process reaches state $v_N$ is
\begin{equation}\label{eq:fixation}
  \frac{1}{1 + \sum_{i = 1}^{N-1} \prod_{j = 1}^i \gamma_j }\,,
\end{equation}
where $\gamma_j$ is the ratio between the probability of going from $v_{j}$ to $v_{j+1}$ and
the probability of going from $v_{j+1}$ to $v_j$.
For the process starting at $v_k$, the probability that the process gets to state $v_N$ is
\begin{equation}\label{eq:fixation_complicated}
  \frac{1 + \sum_{i = 1}^{k-1} \prod_{j = 1}^i \gamma_j }{1 + \sum_{i = 1}^{N-1} \prod_{j = 1}^i \gamma_j }\,.
\end{equation}

We use this equation to argue about a Markov Chain, where the states correspond to the number of mutants on the graph.
We use the equation multiple times for different parts of the spreading of the mutation.

\begin{theorem}\label{thm:environ}
    On a one-dimensional graph (cycle) with segregated configuration, let us denote $r = \frac{f_{A, E}}{f_{B, E}}$ and $r' = \frac{f_{A, \bar{E}}}{f_{B, \bar{E}}}$. 
    \begin{itemize}
    \item If $r,r' > 1$, we have
    \[
        \rho_A = 1 - \frac{1}{2}\left(\frac{1}{r} + \frac{1}{r'}\right) \pm \calO\left(\frac{1}{\sqrt{N}}\right)\,,
    \]
    \item if $r > 1$, $r' < 1$ and $r \cdot r' > 1$, we have
    \[
        \rho_A = \frac{1}{2} - \frac{1}{2r} \pm \calO\left(\frac{1}{\sqrt{N}}\right)\,,
    \]
    \item if $r\cdot r' < 1$, we have
    \[
        \rho_A \le 2^{-\Omega(N)}\,.
    \]
     
    \end{itemize}
\end{theorem}
\begin{proof}
    We begin by considering several degenerate or boundary cases. For some of them, we determine the fixation probability; for some, we argue that we can bound the fixation probability to be $0$ for some starting conditions.
    Moreover, let $E$ be some environment ($R$ or $P$), and $\bar{E}$ the complement to $E$.
    \begin{itemize}
        \item If $f_{A, R} = f_{B, R}$ and $f_{A, P} = f_{B, P}$, we have $\rho_A = \frac{1}{N}$, since the evolution is neutral and one individual eventually wins no matter the configuration.
        \item If $f_{A, E} = f_{B, E}$ and $f_{A,\bar{E}} < f_{B,\bar{E}}$, we have $\rho_A$ exponentially small, since there is a bias against mutants.
        \item If $f_{A, E} = f_{B, E}$ and $f_{A,\bar{E}} > f_{B,\bar{E}}$, we consider the fixation probability of mutant that appears at $E$ to be $0$. This changes the fixation probability by at most $\calO(\frac{1}{N})$.
        \item If $f_{A, E} < f_{B, E}$ and $f_{A,\bar{E}} > f_{B,\bar{E}}$, we consider the fixation probability of mutant that appears at $E$ to be $0$. This again changes the fixation probability by at most $\calO(\frac{1}{N})$.
    \end{itemize}

    First, in environment $E$, where $1 < r = \frac{f_{A, E}}{f_{B, E}}$, we compute the probability that the mutant that appears at a distance at least $2\sqrt{N}$ from the boundary claims at least $\sqrt{N}$ vertices.
    We can treat the graph as a homogeneous isothermal graph on $\sqrt{N}$ vertices.
    Until there are fewer than $\sqrt{N}$ mutants, we can couple the process with the process where the two neighboring residents of mutants are connected in a cycle.
    Then, the transition probabilities are the same in this cycle and on our part of the path.
    The only exception happens when the mutants spread $2\sqrt{N}$ steps towards the boundary without claiming $\sqrt{N}$ vertices.
    Note that the reproduction or death of a mutant happens with the same probability, either closer to the boundary or further away from the boundary.
    This means we have the probability that the mutant closer to the boundary experiences by $2\sqrt{N}$ more reproductions than deaths, and the mutant further away from the boundary experiences $\sqrt{N}$ more deaths than reproductions is at most $2^{-\sqrt{N}}$.
    The probability that one mutant spreads to at least $\sqrt{N}$ vertices is proportional to
    \[
        \frac{1-\frac{1}{r}}{1-\frac{1}{r^{\sqrt{N}}}}\,,
    \]
    with additive error proportional to $2^{-\Omega(\sqrt{N})}$ from the isothermal theorem.

    Second, we compute the probability that the mutants spread over the rest of the half of the graph.
    For $r>1$, this means from \Cref{eq:fixation} that the process needs to make $\sqrt{N}$ more negative steps (deaths) than positive (reproductions) against the bias.
    This has the probability at most $2^{-\Omega(\sqrt{N})}$.

    Finally, we compute the probability that the mutants that occupy the environment $E$ fully spread to the environment $\bar{E}$.
    If the fitness of mutants in both environments is higher, this happens with probability close to $1$.

    So in this case, we consider $1 > r' = \frac{f_{A, \bar{E}}}{f_{B, \bar{E}}}$.
    Let the $N$ vertices of the cycle be numbered clockwise, and the first $N/2$ vertices are from environment $E$.
    At the beginning, mutants occupy vertices $1,2,\dots,N/2$ and residents occupy vertices $N/2+1, \dots N$.
    There are two boundaries between mutants and residents, one of them is between $1$ and $N$, and the other between $N/2$ and $N/2+1$.
    These boundaries move randomly and independently, given by the fitness ratios $r$ and $r'$.
    We examine only one boundary, without loss of generality, between $N/2$ and $N/2+1$ and freeze the other; mutants win when they spread to $N$, and the residents win if they spread to $1$. Since environments have only a simple structure of heterogeneity, it is a good estimate of the fixation probability.

    The evolution of mutants on this graph can be viewed as a random walk on the Markov Chain, where we are interested only in how the boundary moves.
    For the probability computation, we use \Cref{eq:fixation_complicated},
    where $\gamma_j = \frac{1}{r}$ for $j \in \{1,N/2\}$ and $\gamma_j = \frac{1}{r'}$ for $j \in \{N/2+1,N\}$.
    Plugging into \Cref{eq:fixation_complicated}, we obtain
    \[
        \frac{1 + \sum_{i = 1}^{N/2}  \frac{1}{r^i} }{1 + \sum_{i = 1}^{N/2}  \frac{1}{r^i} + \frac{1}{r^{N/2}} \cdot \sum_{i = N/2+1}^{N-1} \frac{1}{(r')^i} }\,.
    \]
    We bound the value of $\frac{1}{r^{N/2}} \cdot \sum_{i = N/2+1}^{N-1} \frac{1}{(r')^{i-N/2}}$.
    This is proportional to $\left(\frac{1}{r\cdot r'}\right)^{\frac{N}{2}}$.
    This gives us three possible scenarios for the magnitude of the expression $\frac{1 + \sum_{i = 1}^{N/2}  \frac{1}{r^i} }{1 + \sum_{i = 1}^{N/2}  \frac{1}{r^i} + \frac{1}{r^{N/2}} \cdot \sum_{i = N/2+1}^{N-1} \frac{1}{(r')^i} }$.
    For $r \cdot r' > 1$, the value is $\frac{1}{1 + 2^{-\Omega(N)}}$.
    For $r\cdot r' = 1$, the fixation probability of mutants is proportional to $\frac{1}{2}$.
    For $r \cdot r' < 1$, the value is $\frac{1}{1 + 2^{\Omega(N)}}$.
   
\end{proof}

\newpage

\section*{Supplementary Note 5: Interleaved Configuration}\label{sec:app_interleaved}

In this section, we examine a one-dimensional configuration in which two vertices in a rich environment alternate with two vertices in a poor environment. We refer to this graph as a $2$–$2$ cycle.

For Scenario $1$ (where both types are affected by the environment), we compute the probability that a new mutant dies in the first step. This scenario is the only case in which the interleaved configuration produces a non-monotonic dependence on spatial mixing.

Then, we show that for some parameters of mutant advantage and $\sigma$, the fixation probability is smaller than in the separated environment and in the completely mixed environment.

\begin{lemma}\label{lem:interleaved}
    Given mutant fitness $r$ and deviation $\sigma$, in Scenario 1 (S1), the probability that the mutant dies before it reproduces is
    \[
        \frac{r+1}{(r+1)^2 - \sigma^2}\,.
    \]
\end{lemma}

\begin{proof}
    There are only two non-isomorphic positions where the mutant can appear, either at a poor vertex or a rich vertex.
    In both cases, the new mutant has two resident neighbors in different environments.
    The mutant dies if the neighboring resident is selected and then chooses the mutant to be replaced (it has two neighbors, which occurs with probability $\frac{1}{2}$).
    The mutant reproduces if it is selected.
    We can examine only these active steps.

    If the mutant appears in a rich environment, the probability of death before the first reproduction is
    \[
        \frac{\frac{1}{2}(1 + \sigma) + \frac{1}{2}(1 - \sigma)}{r + \sigma + \frac{1}{2}(1 + \sigma) + \frac{1}{2}(1 - \sigma)} = \frac{1}{r + \sigma + 1}\,.
    \]
    Similarly, if the mutant appears in a poor environment, the probability of death before reproduction is
    \[
        \frac{1}{r - \sigma + 1}\,.
    \]

    Averaging both over the starting positions gives
    \[
        \frac{1}{2}\frac{1}{r + \sigma + 1} + \frac{1}{2}\frac{1}{r - \sigma + 1} = \frac{r+1}{(r+1)^2 - \sigma^2}\,.
    \]
\end{proof}

Now, we formalize the observation from Figure 4 in the main text.

\begin{theorem}
    In Scenario~1, for $r \ge 1.2$ and $\sigma \ge 0.95$, the $2$–$2$ cycle graph has a lower fixation probability than the segregated configuration or the checkerboard distribution.
\end{theorem}

\begin{proof}
    From \Cref{lem:interleaved}, we have, the fixation probability is at most 
    \[
        1- \frac{r+1}{(r+1)^2 - \sigma^2} = \frac{(r+1)^2 - \sigma^2 - (r+1)}{(r+1)^2 - \sigma^2} = \frac{r(r+1)-\sigma^2}{(r+1)^2 - \sigma^2}
    \]
    for the $2-2$ cycle.
    Note that this is a crude estimate that takes into account only the first step of the process.
    
       From \Cref{thm:environ}, we have that the fixation probability is $1 - \frac{1}{2}\left(\frac{1+\sigma}{r+\sigma} + \frac{1-\sigma}{r-\sigma}\right) = 1-\frac{r-\sigma^2}{r^2-\sigma^2}$ for the segregated environment.
    From Supplementary Note~3, Scenario~1, we have that the fixation probability is, again, $1-\frac{r-\sigma^2}{r^2-\sigma^2}$ for the checkerboard environment.

    This implies that for $r \ge 1.2$ and $\sigma \ge 0.95$, the $2$–$2$ interleaved configuration yields a lower fixation probability than either the segregated or checkerboard configuration. These parameter values correspond to the regime illustrated in Fig.~4.

\end{proof}

\newpage

\section*{Supplementary Note 6: Overview of previous findings in the literature}

In this section, we review earlier theoretical studies of fixation and invasion
dynamics in heterogeneous environments and place them within the framework
developed in the main text, emphasizing the roles of genotype specificity and
spatial arrangement.\\

\noindent
\textbf{Metapopulation Island models.}   
Early works on evolution in heterogeneous environments focused on geographically
subdivided populations (demes or islands). Analytical results for the fixation probability within the diffusion-approximation regime were reported. \cite{Gavrilatzlets2002fixation, whitlock2005probability, tachida1991fixation}.
In these models, populations occupy a small number of well-mixed habitats connected through migration,
and environmental heterogeneity enters through habitat-dependent selection
coefficients.

Under weak selection, the fitness difference between mutants and residents in deme
$i$ is summarized by a single local selection coefficient $s_i (= f_{\rm A,i} - f_{\rm B,i})$. Because diffusion
approximations depend only on these effective selection coefficients, they cannot
distinguish whether environmental effects act asymmetrically on both genotypes or
preferentially on one genotype, provided the net selection coefficient is unchanged.
Consequently, the genotype specificity of environmental effects is not resolvable within
this framework.

Across different migration regimes, these studies reported that
environmental heterogeneity increases the fixation probability whenever mutants are
beneficial on average. In our framework, this corresponds to the joint weak-selection
and weak-heterogeneity limit of segregated configuration, where mutant-specific (Scenario~2) and resident-specific
(Scenario~3) environments become effectively indistinguishable. Also, as discussed in the main text, this regime of weak-heterogeneity and weak-selection for segregated configuration, represents the only outlier in our claim that a mutant-specific environment suppresses selection.\\

\noindent
\textbf{Evolutionary graphs with location-dependent fitness.} 
A second class of studies considers Moran processes on explicit graphs with
location-dependent fitness, resolving discrete individuals, finite population size,
and spatial structure beyond diffusion approximations.

On complete graphs, Hauser and Nowak \cite{hauser2014heterogeneity} studied symmetric
fitness heterogeneity affecting both genotypes and found that heterogeneity suppresses
selection. Later, Kaveh et al. \cite{kaveh2019environmental} proved that, for sufficiently large
complete graphs, the fixation probability depends only on mutant fitness heterogeneity and
is always suppressed by it, whereas heterogeneity in resident fitness can produce a
weak amplification for smaller population sizes while being irrelevant for larger population sizes.

Related effects have been reported on lattice graphs and cycles with quenched or random
fitness landscapes \cite{mahdipour2017genotype, farhang2017effect,farhang2019environmental}. 
In these studies, fitness values are typically drawn independently across locations, so mutants and 
residents experience distinct local fitness values despite identical distributions on average. This setup
corresponds most closely to genotype-symmetric environments (Scenario~1) combined with
random or weakly correlated environmental configurations. In this regime, amplification
arises only through second-order spatial effects, consistent with the modest amplifications
observed in the main text. Similarly, \cite{manem2015modeling} analyzed a random fitness distribution model on a lattice where heterogeneity mainly affects mutants, and the authors observed a decrease in the fixation probability. There are other works in the literature \cite{giaimo2018invasion, maciejewski2014environmental} that used node-dependent fitness models, but the analysis is often brief.\\

\noindent
\textbf{Environmental isothermal theorem.}  
A unifying perspective emerges from results on bipartite graphs with
environment-dependent fitness \cite{kaveh2020moran}. For such graphs, the fixation
probability is independent of degree connectivity and depends only on the fitness
values associated with the two partitions. This ``environmental isothermal theorem''
implies that checkerboard arrangements on lattices, complete bipartite graphs, or any bipartite or properly two-colorable regular graphs, such as square-lattice, hexagonal lattice, and cycle graph, are evolutionarily equivalent. 

In the context of the present work, this result explains why checkerboard environments admit exact
analytical solutions and why fixation probabilities in these environments depend only
on genotype–environment interaction scenarios, not spatial dimension or local
connectivity. More broadly, bipartite graphs provide a bridge between classical
population-genetic models and spatial evolutionary graph theory, clarifying which
effects of environmental heterogeneity are universal and which depend on spatial
organization. These results place bipartite and high-migration models squarely in the
checkerboard regime summarized in Fig.~5, where genotype specificity determines the
direction of selection.

\section*{Supplementary Note 7: Supplementary Figures and Tables}\label{sec:figures_and_tables}

\begin{figure}
\begin{center}
\captionsetup{justification=raggedright,singlelinecheck=false}
\includegraphics[width=.8\textwidth]{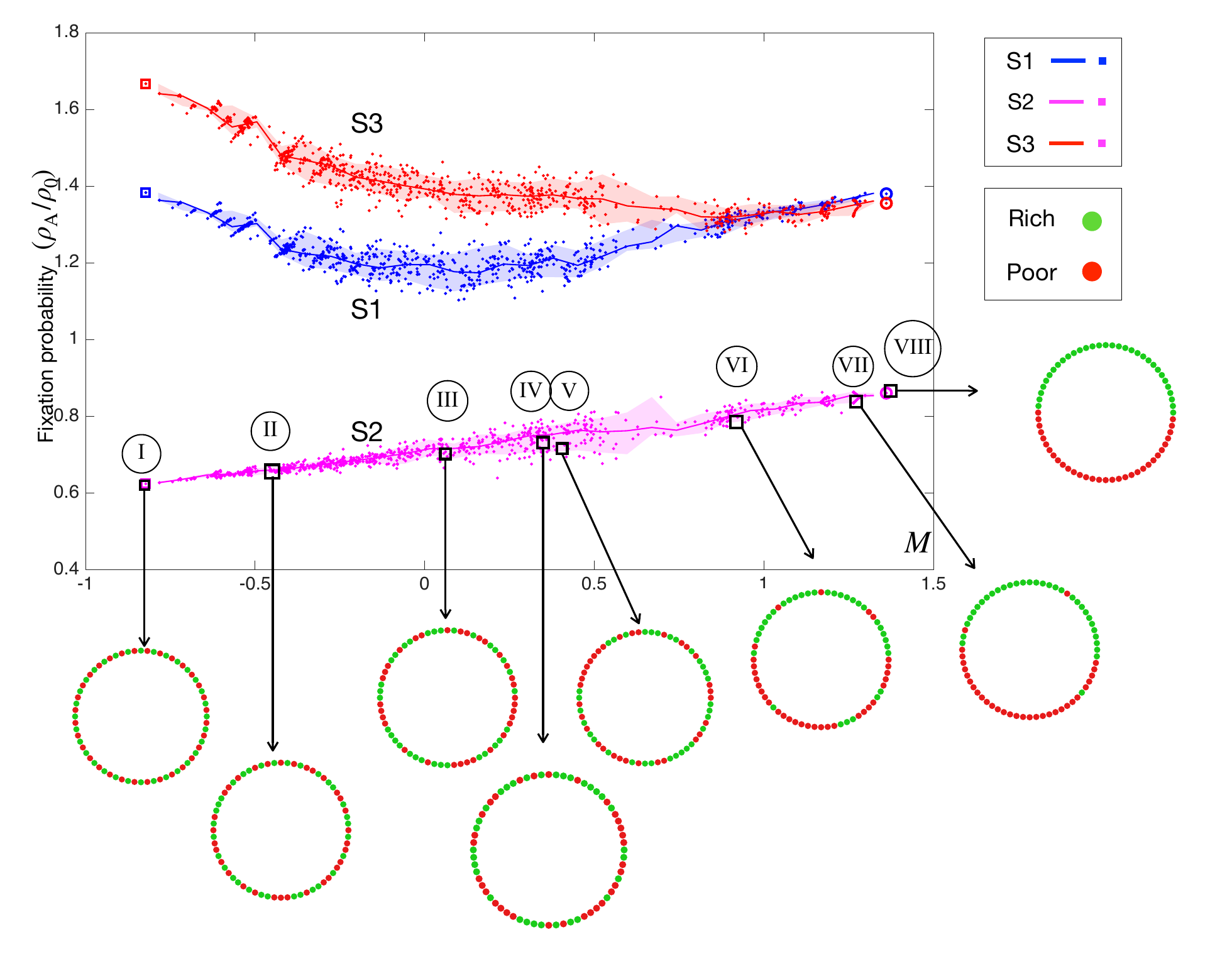}
\caption{\textbf{Fixation probability versus environmental correlation index for representative configurations.}
Fixation probability $\rho_{\rm A}$ as a function of the spatial correlation index $M$ for a one-dimensional cycle ($N=64$) at fixed heterogeneity amplitude $\sigma=0.8$ and selection strength $r=1.5$. Each point corresponds to a distinct environmental configuration, sampled uniformly across the full range of $M$, with the same configuration shown separately for each interaction scenario (S1–S3). Colors denote the three genotype–environment interaction scenarios, as defined in Fig.~ 1.  Selected configurations are shown explicitly to illustrate how different values of $M$ correspond to qualitatively distinct spatial arrangements, ranging from highly intermixed (checkerboard-like) to strongly segregated domains. This figure complements Fig.~4 in the main text by visualizing the geometric structure underlying the trend for the fixation probabilities.
}
\label{figS1}
\end{center}
\end{figure} 

\begin{figure}
\begin{center}
\captionsetup{justification=raggedright,singlelinecheck=false}
\includegraphics[width=.8\textwidth]{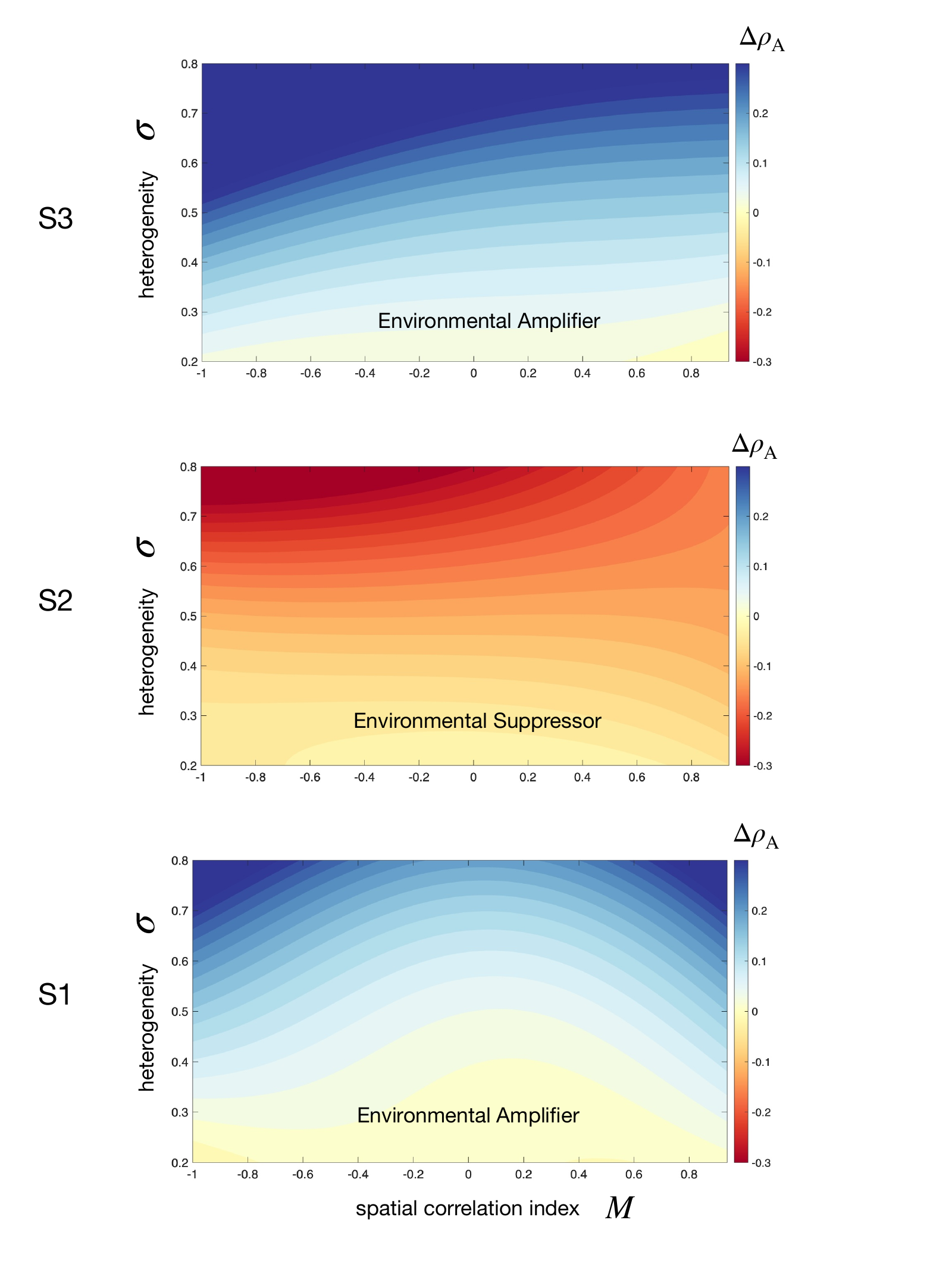}
\caption{\textbf{Smoothed fixation probability landscape in the $(M,\sigma)$ plane.}
Heat map showing the deviation of fixation probability, $\Delta\rho_{\rm A} = \rho_{\rm A}-\rho_{\rm A}(\sigma=0)$, as a function of the spatial correlation index $M$ and heterogeneity amplitude $\sigma$ for a one-dimensional cycle ($N=64$, $r=1.5$). Fixation probabilities were computed for discrete values $\sigma=0.2,0.4,0.6,0.8$ and 900 environmental configurations at each $(M,\sigma)$. The heat map and contours are obtained by polynomial interpolation and smoothing of these data to guide the eye. Separate panels correspond to the three genotype–environment interaction scenarios (S1–S3).
This figure provides a complementary visualization of Fig.~4 in the main text, illustrating how spatial environmental arrangement and heterogeneity jointly shape amplification and suppression regimes.
}
\label{figS2}
\end{center}
\end{figure} 

\begin{figure}
\begin{center}
\captionsetup{justification=raggedright,singlelinecheck=false}
\includegraphics[width=1\textwidth]{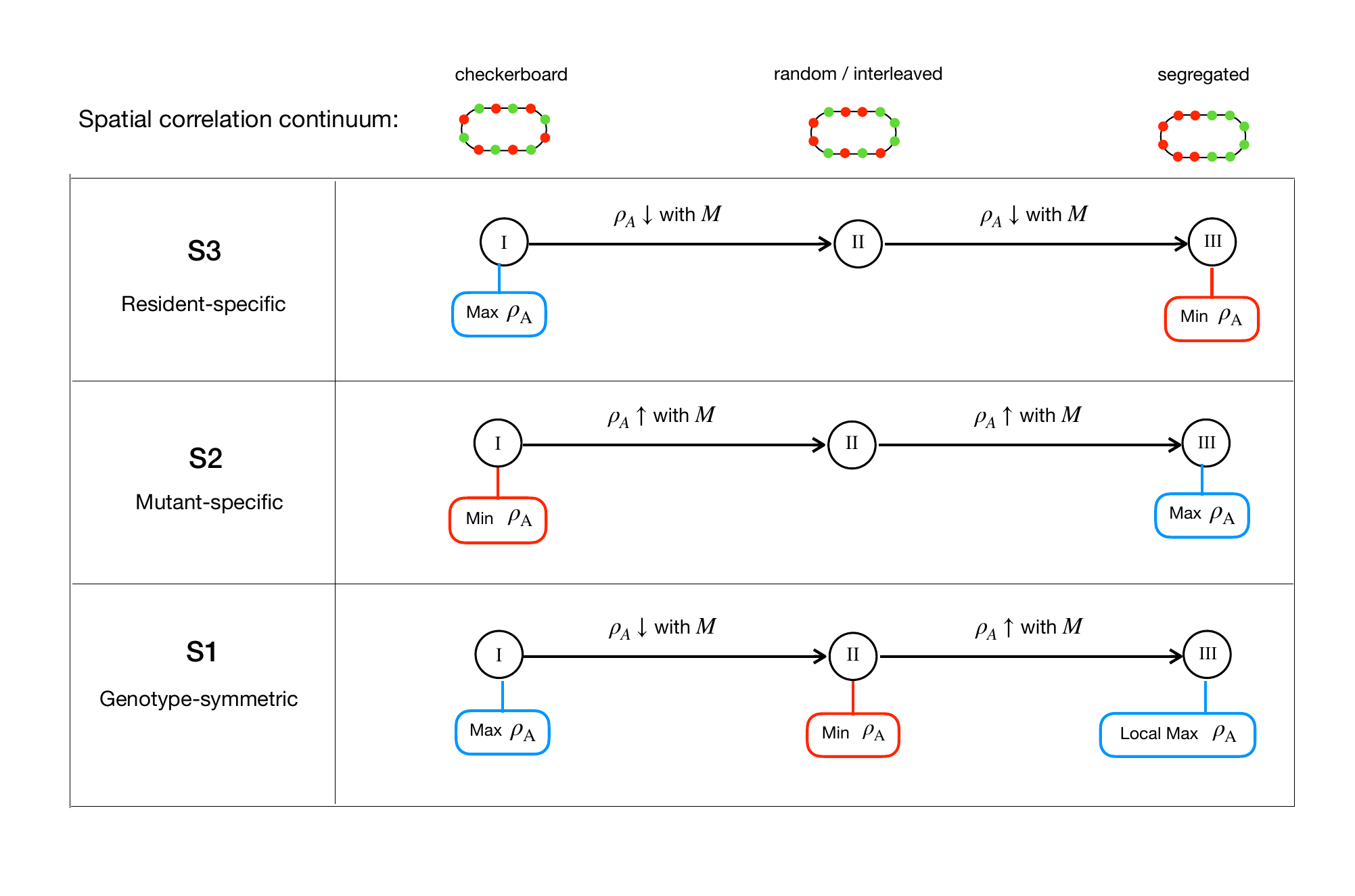}
\caption{\textbf{Schematic locations of the extrema in the fixation probability.}
Schematic summary of where the maxima and minima of the fixation probability $\rho_{\rm A}$ along the spectrum of $M$, for the three genotype–environment interaction scenarios ($r=1.5$, $\sigma = 0.8$). Arrows indicate qualitative trends with increasing spatial correlation index ($M$). For resident-specific environments (Scenario~3), fixation is maximized in highly intermixed configurations and decreases with clustering. For mutant-specific environments (Scenario~2), the trend reverses, with maximal fixation in clustered landscapes. For genotype-symmetric environments (Scenario~1) fixation probability is non-monotonic in $M$, attaining a minimum at intermediate values of $M$. This schematic summarizes the locations of extrema observed in Fig.~4 for Scenarios~S1--S3.
}
\label{figS4}
\end{center}
\end{figure} 

\begin{figure}
\begin{center}
\captionsetup{justification=raggedright,singlelinecheck=false}
\includegraphics[width=1\textwidth]{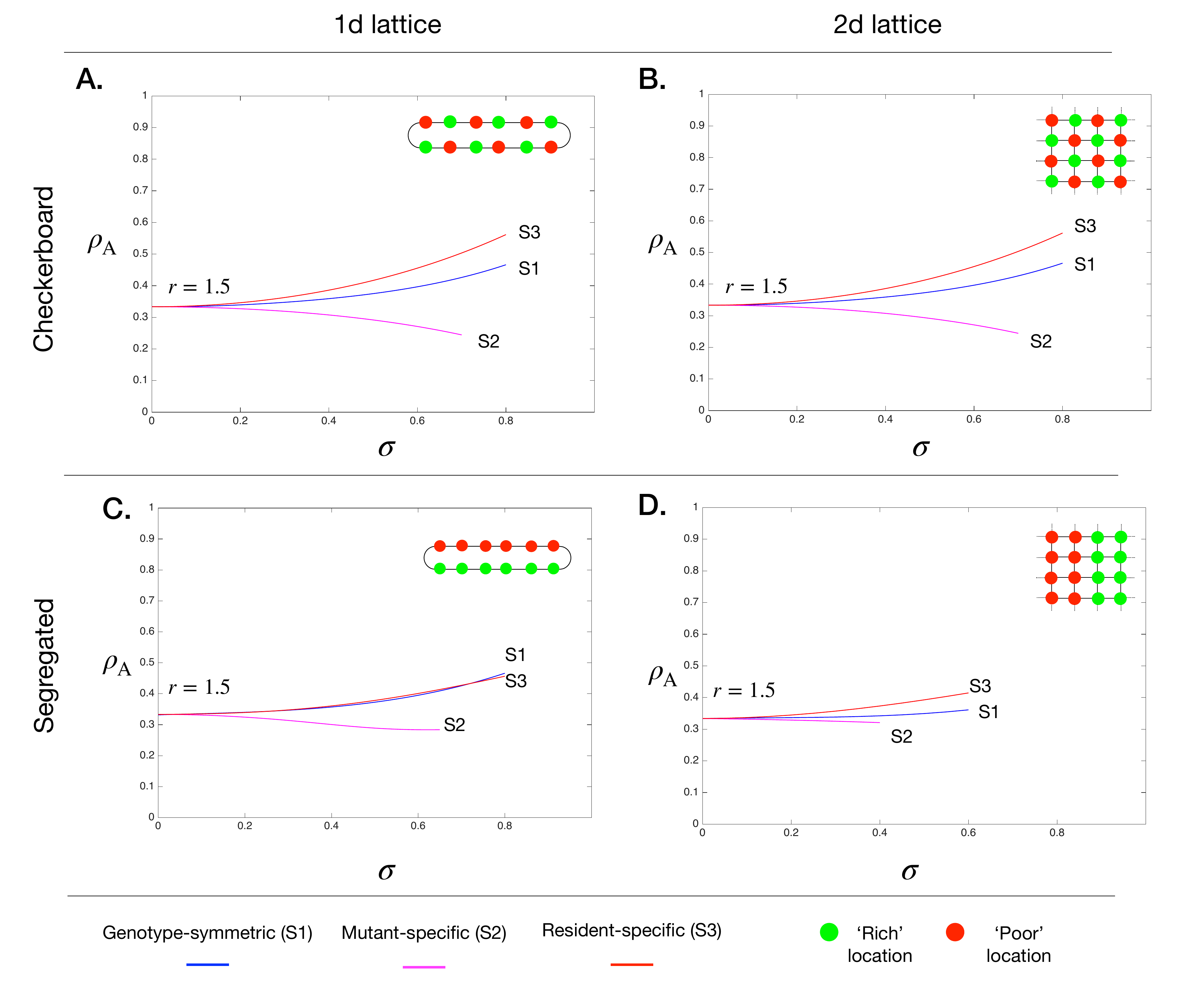}
\caption{\textbf{Fixation probability as a function of heterogeneity amplitude $r=1.5$.}
Fixation probability $\rho_{\rm A}$ as a function of $\sigma$ for the three
genotype–environment interaction scenarios (S1–S3). Panels correspond to:
(\textbf{A}) 1D cycle, checkerboard (maximally intermixed);
(\textbf{B}) 2D lattice, checkerboard (maximally intermixed);
(\textbf{C}) 1D cycle, segregated (highly clustered);
(\textbf{D}) 2D lattice, segregated (highly clustered).
Each panel shows results for $r=1.5$ to complement the results reported in Fig.3 (main text) Solid curves denote numerical or simulation results; in checkerboard environments, they coincide with the analytical prediction. Population size $N=64$.}
\label{figS3}
\end{center}
\end{figure} 

\begin{table}
\centering
\includegraphics[width= 1\textwidth]{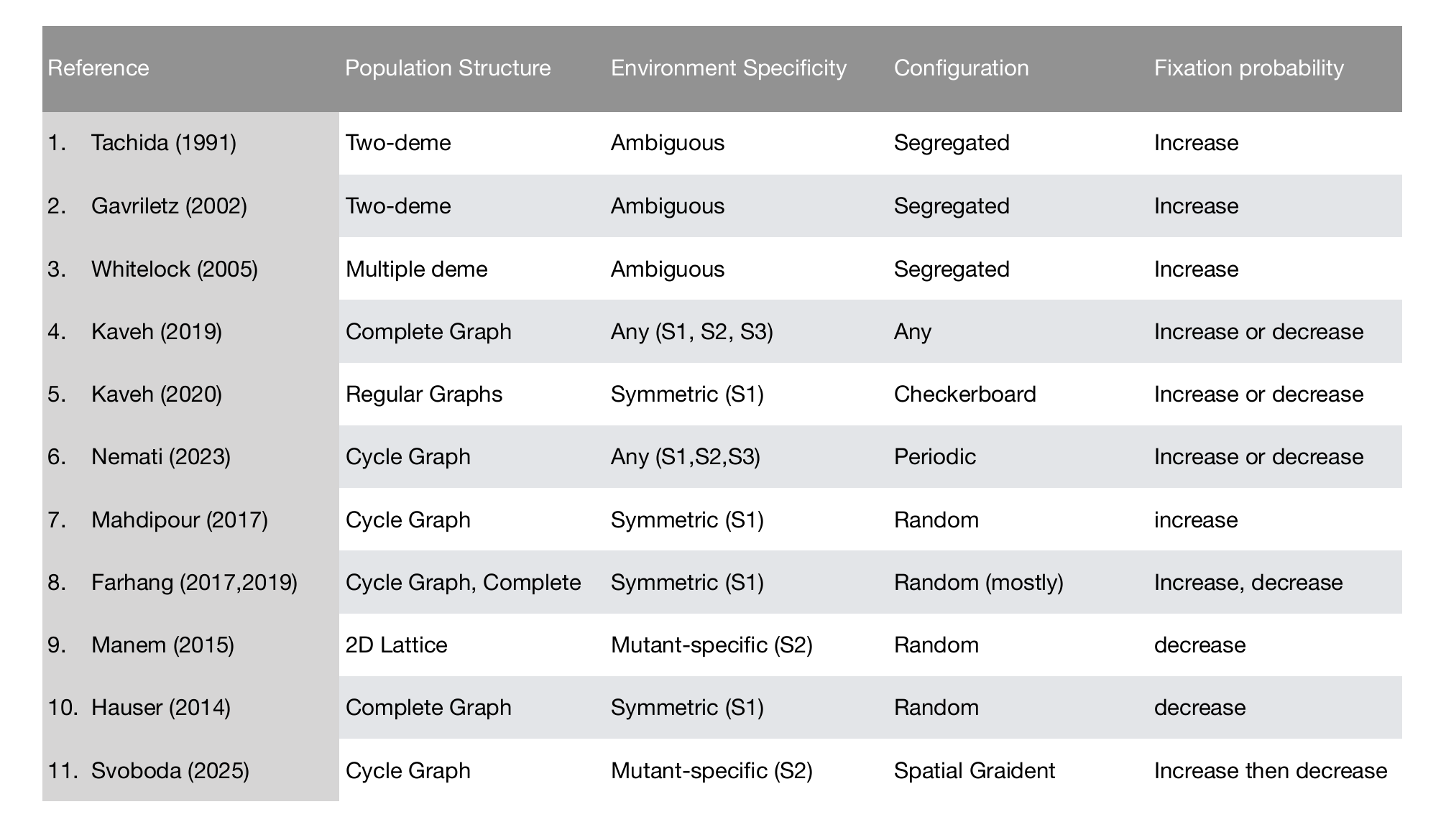}
\caption{\textbf{Summary of theoretical results on fixation in heterogeneous environments.}
Representative studies on evolutionary dynamics in spatially or environmentally heterogeneous systems, categorized by population structure, type of fitness heterogeneity, and reported effect on fixation probability. The final column indicates how each study maps onto the present framework in terms of genotype specificity (mutant-specific, resident-specific, or genotype-symmetric) and spatial environmental arrangement. This table with results in Fig.5 shows that seemingly contradictory amplification or suppression effects reported in the literature are consistent once genotype specificity and environmental arrangement are made explicit.
}

\label{tab:S1}
\end{table}

\begin{table}
\centering
\includegraphics[width=0.8\textwidth]{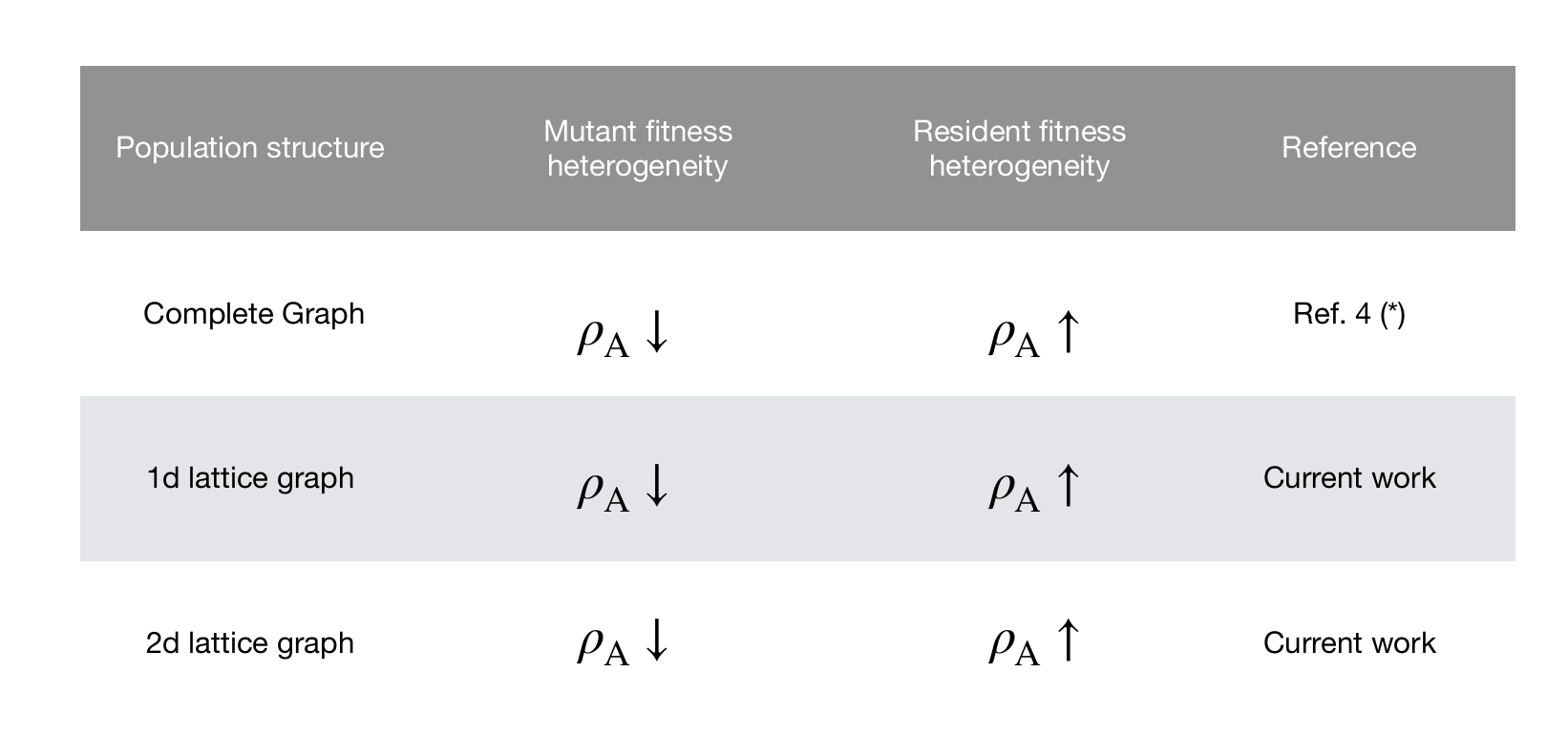}
\caption{\textbf{Effect of graph structure and connectivity on fixation probability in heterogeneous environments.}
Summary of fixation behavior across different graph classes, highlighting how genotype-specific environmental heterogeneity interacts with spatial structure. While the direction of amplification or suppression is primarily determined by genotype specificity, the influence of spatial correlation, $M$, diminishes as graph connectivity increases. In highly connected graphs, spatial correlations become irrelevant, and fixation outcomes depend mainly on the distribution of fitness values. ($*$: In Ref.~4 the increase in the fixation probability due to the resident fitness heterogeneity is only true for small complete graphs, $N \lesssim 10$.)
}
\label{tab:S2}
\end{table}

\clearpage
\bibliographystyle{unsrtnat}
\bibliography{references}